%% file: bttqst_main.tex
\documentclass[final]{siamart251216}
\usepackage{graphicx}%
\usepackage{mathrsfs}%
\usepackage[title]{appendix}%
\usepackage{textcomp}%
\usepackage{manyfoot}%
\usepackage{float}
\usepackage{tabularx, booktabs, multirow, siunitx, colortbl}
\usepackage{caption}
\usepackage{subcaption}
\usepackage[normalem]{ulem}
\usepackage{wrapfig}
\usepackage{pgfplots}
\usetikzlibrary{pgfplots.groupplots, pgfplots.statistics}
\usepgfplotslibrary{fillbetween}
\usepackage{pgfplotstable}
\pgfplotsset{compat=1.18} 
\usepackage{comment}
\usepackage[most]{tcolorbox}

\usepackage{algorithmic}

\newcommand{\ttc}[1]{%
\textcolor{green!50!black}{\footnotesize\texttt{/* #1 */}}%
}

\usepackage{soul}
\usepackage{xcolor}
\sethlcolor{yellow}

\usepackage{extramath}
\newcommand{\rank}[1]{\text{rank}\left( #1 \right)} 

\newcommand{\ubar}[1]{\underline{\smash{#1}}}

\let\oldrho\rho
\renewcommand{\rho}{\bm{\oldrho}}

\newcommand{\sktensor}{\mathbin{\lvert \!\otimes\! \rvert}}
\newcommand{\sbullet}{\mathbin{\lvert\, \text{\tiny$\bullet$}\,\rvert}}

\newcommand{\ttm}[1]{\mat{#1}_{\scriptscriptstyle\mathsf{TT}}}

\newcommand{\ttmm}[1]{\mat{#1}_{\scriptscriptstyle\mathsf{TT}, {\scriptstyle m}}}

\newcommand{\eff}[2]{\mat{#1}^{\scriptstyle (\!#2\!)}_{\scriptstyle\mathsf{eff}}}

\newcommand{\effm}[2]{\mat{#1}^{\scriptstyle (\!#2\!)}_{\scriptstyle\mathsf{eff}, m}}

\newcommand{\matL}[1]{\mat{#1}_{\scriptscriptstyle\mathsf{L}}}
\newcommand{\matR}[1]{\mat{#1}_{\scriptscriptstyle\mathsf{R}}}

\definecolor{virpurple}{RGB}{68, 1, 84}   
\definecolor{virblue}{RGB}{49, 104, 142}   
\definecolor{virgreen}{RGB}{53, 183, 121}  
\definecolor{viryellow}{RGB}{194, 184, 9}

\mathtoolsset{showonlyrefs=false}

\title{A Block Tensor Train Burer-Monteiro Framework for Low-Rank Quantum State Tomography\thanks{This paper is an extended version of our conference paper {\em Tensor Train Quantum State Tomography using Compressed Sensing},  in the 33rd European Signal Processing Conference (EUSIPCO), pp. 1332–1336, 2025. 
See the contribution statement for details on new additions.
\funding{This work was supported by the Flemish Government's AI Research Program and KU Leuven Internal Funds (iBOF/23/064, C14/22/096). Shakir Showkat Sofi (\email{shakirshowkat.sofi@kuleuven.be}), Charlotte Vermeylen (\email{charlotte.vermeylen@kuleuven.be}), and Lieven De Lathauwer (\email{lieven.delathauwer@kuleuven.be}) are affiliated with Leuven.AI - KU Leuven institute for AI, B-3000, Leuven, Belgium.}}}

\author{
Shakir Showkat Sofi\thanks{
Department of Electrical Engineering (ESAT), KU Leuven, Leuven, Belgium;
Group Science, Engineering and Technology, KU Leuven Kulak, Kortrijk, Belgium.
}
\and
Charlotte Vermeylen\footnotemark[2]
\and
Fatemeh Mohammadi\thanks{ Departments of Computer Science and Mathematics, KU Leuven, Leuven, Belgium.\\
(\email{fatemeh.mohammadi@kuleuven.be}).}
\and
Lieven De Lathauwer\footnotemark[2]
}
\begin{document}
\maketitle

\begin{abstract}
Quantum state tomography is a fundamental technique for estimating the state of a quantum system from measured data and plays a crucial role in evaluating the performance of quantum devices. However, standard estimation methods become computationally prohibitive as the system size increases due to the exponential growth of the density matrix, describing a quantum state, with the number of qubits. We propose a low-rank tensor-network framework for mixed-state quantum state tomography based on a block tensor train (Block-TT) factorization. Specifically, the density matrix is represented as the contraction of a Block-TT with its Hermitian transpose, yielding a TT analogue of the Burer-Monteiro factorization. This parameterization guarantees Hermiticity and positive semidefiniteness by construction while compressing the number of optimization variables from exponential to linear in the number of qubits. Building on this representation, we develop single-site and two-site density matrix renormalization group (DMRG) algorithms for estimating quantum states from compressed measurements. The resulting methods operate directly on the compressed parameterization, support adaptive rank refinement, and exploit efficient tensor-network contractions for expectation-value evaluation. The framework is applicable to a broad class of low-rank quantum states, including pure states, nearly pure states, and ground states that admit accurate tensor-network approximations. Numerical experiments demonstrate accurate state reconstruction from limited measurements together with substantial reductions in memory requirements and computational cost compared with conventional low-rank tomography methods.
\end{abstract}

\begin{keywords}
tensor networks, low-rank approximation,  tensor trains, compressed sensing, quantum state tomography
\end{keywords}

\begin{AMS}
81P18, 15A69, 65F55
\end{AMS}

\pagestyle{myheadings}
\thispagestyle{plain}

\section{Introduction}
Quantum computing has attracted significant interest in recent years, driven by emerging tools for analyzing large-scale systems and optimizing over high-dimensional spaces. The study of quantum systems is increasingly important in both academia and industry, with applications spanning quantum communication, sensing, and control. A quantum system’s state is fully described by its density matrix—a Hermitian, positive semidefinite (PSD) operator with unit trace. It provides a unified representation of both pure states and mixed states (probabilistic mixtures of pure states), where diagonal entries correspond to state populations and off-diagonal entries encode quantum coherences \cite{paris2004qse, nielsen2010quantum}.

A typical quantum algorithm begins by initializing a system in a specific quantum state (state preparation), followed by manipulating the state via quantum gates (state evolution), and finally extracting classical information through measurement (state readout/tomography). Each stage presents its own challenges and, in particular, the final step---quantum state tomography (QST)---is fundamentally challenging. QST is an inverse problem: from the measurements, the state cannot be accessed directly; instead, the measurements yield samples from a probability distribution determined by the quantum state. The goal of QST is to reconstruct the quantum state from these samples.
Two key challenges arise in this reconstruction problem. First, the number of measurements—and the associated computational cost—required to reconstruct a generic, unstructured state grows exponentially with the number of qubits defining it, an obstacle known as the  ``curse of dimensionality." Second, the reconstructed state must satisfy the physical constraints of quantum mechanics, namely Hermiticity, positive semidefiniteness, and unit trace, while remaining amenable to efficient storage and post-processing, such as the computation of expectation values. This naturally raises a fundamental question: which quantum states admit efficient representations, and what structural properties enable them?  In practice, many quantum states of interest exhibit special structural properties. These include pure or low-entropy states, ground states of local Hamiltonians, product states, and other structured states in which only a subset of entries contains significant information  \cite{eisert2010colloquium, gross2010csqst, liu2012csqst}. Many of these examples can be unified under the broader framework of data-sparse or low-rank (in the matrix or tensor sense) structures. QST for such structured states can be performed efficiently by leveraging this structure, as only a partial set of measurements may be sufficient for reconstruction \cite{gross2010csqst, liu2012csqst}. \par

\subsection{Related Work} \label{subsec:overview}
When the density matrix is known to be (approximately) low rank, \emph{low-rank} QST exploits this structure to reconstruct the state from significantly fewer measurements. Rank constraints can be incorporated either implicitly, by promoting low rank via a rank-minimization objective implemented through the nuclear norm as a convex surrogate under measurement constraints \cite{gross2010csqst, liu2011universal, liu2012csqst, wang2013qstvmc}, or explicitly, by optimizing a fixed-rank factorized model, as in Burer--Monteiro (Cholesky-like) formulations \cite{burer2003lrsdp, kyrillidis2018provable}.  Nevertheless, the dependence of the model parameters remains exponential in the number of qubits. 

Alternatively, when the density matrix is represented as a high-order tensor, tensor networks---interconnected networks of low-order tensors---are employed to achieve compact representations and alleviate the curse of dimensionality. Prominent tensor network formats include matrix product states (MPS, also known as tensor trains (TT)), tree tensor networks, and projected entangled pair states \cite{verstraete2007MPS, oseledets2010tensortrain, orus2014practical, khoromskaia2018tensor}. In particular, MPS comes with highly efficient optimization schemes based on the density matrix renormalization group (DMRG) \cite{white1993density,schollwock2005dmrg, white2005density, khoromskij2010dmrgqtt,  oseledets2011dmrg, holtz2012alternating, rohwedder2013ttopt, kressner2014btt, lars2015ttals}, making \emph{MPS-based} QST scalable for large-scale quantum many-body systems whose states admit low-rank tensor network representations, including ground states, GHZ states, cluster states, and AKLT states \cite{cramer2010efficient, lanyon2017efficient, kuzmin2024qst, sofi2025bttqst}.  In \cite{qin2024quantum}, QST is generalized to matrix product operators (MPO), extending MPS methods from pure states to noisy mixed states \cite{verstraete2004MPO}.  A key challenge in MPO-based approaches is ensuring that the reconstructed density matrix remains physically valid \cite{verstraete2004MPO, cuevas2013purifications}. Although specialized variants with constraints on MPO cores can enforce positivity \cite{cuevas2013purifications}, designing DMRG-like algorithms in this setting is generally challenging, and such algorithms remain underexplored for the QST task. While there are a few works in this direction, they do not address the original QST problem; instead, they focus on estimating the underlying density matrix from reduced density matrices \cite{cramer2010efficient, guo2024quantum}.

A complementary line of work avoids full state reconstruction and instead targets specific structural or functional aspects of the density matrix. For instance, \emph{permutationally invariant} QST efficiently reconstructs the permutationally invariant component of a state, with measurement complexity scaling only quadratically in the number of qubits, and is particularly well suited for states close to permutation symmetry, such as Dicke and spin-squeezed states \cite{toth2010permutationally, moroder2012permutationally}. In a different direction, \emph{shadow tomography} constructs compact classical representations (classical shadows) via randomized measurements, enabling efficient estimation of many observables with sample complexity that scales only logarithmically in the number of observables \cite{aaronson2018shadow, huang2020predicting}. Extending this idea of partial information extraction, \emph{selective} QST focuses on estimating only a chosen subset of density matrix entries, avoiding full reconstruction \cite{baldwin2016strictly, calderaro2018direct, feng2021direct, morris2019selective, sofi2026tomography}. One of the questions in this class of methods is whether such partial information is sufficient to infer reliable information about the full state, or, in other words, whether it is possible to use such methods to guarantee the recoverability of the full state, which is of actual interest in many applications. An algebraic completion-based approach with deterministic recovery conditions leveraging selective measurements has been developed in \cite{sofi2026tomography}

\subsection{Contributions} In this work, we present a block tensor train (Block-TT) Burer-Monteiro framework for learning high-dimensional mixed quantum states via compressed sensing. By compressing the parameter space from exponential to linear in the number of qubits (or qudits, more generally), the Block-TT Burer-Monteiro representation directly mitigates the curse of dimensionality while inherently ensuring physical state validity by construction. Extending our preliminary conference work \cite{sofi2025bttqst}, where this representation was first introduced, this paper makes the following contributions:
\begin{itemize}
	\item We develop novel single-site and two-site DMRG algorithms for compressed-sensing quantum state tomography in Block-TT format. The proposed methods operate directly on the low-rank factor representation, preserve positivity by construction, and support adaptive rank refinement through tensor-network sweeps.
	\item We derive efficient tensor-network contraction schemes based on effective operators and recursively constructed environments, enabling scalable expected-value evaluation and local optimization.
    \item We extend the original conference framework with overlapping measurement operators and environment caching strategies, allowing efficient tomography of larger structured quantum systems.
    \item We provide an extensive numerical study comparing the proposed approaches with established low-rank tomography methods, demonstrating improved reconstruction accuracy, rank adaptivity, and computational efficiency across a range of problem settings.
\end{itemize}

\subsection{Outline}
\Cref{subsec:prelims} reviews the fundamentals of QST and TT representations, introducing the definitions used throughout the paper.  \Cref{sec:bttqst} presents our main contributions, including a compressed sensing approach for low-rank QST in the proposed block-TT format in \Cref{subsec:csqst}. \Cref{sec:exps} presents supporting numerical experiments and \Cref{sec:con} concludes the paper.

\section{Preliminaries}\label{subsec:prelims}

This section establishes the notation and reviews the concepts from low-rank QST and TT representations that are needed for the derivation of the Block-TT tomography algorithms in \Cref{sec:bttqst}.

\subsection{Notation}
\begin{figure}[htb]
	\centering
	\includegraphics[width=0.85\linewidth]{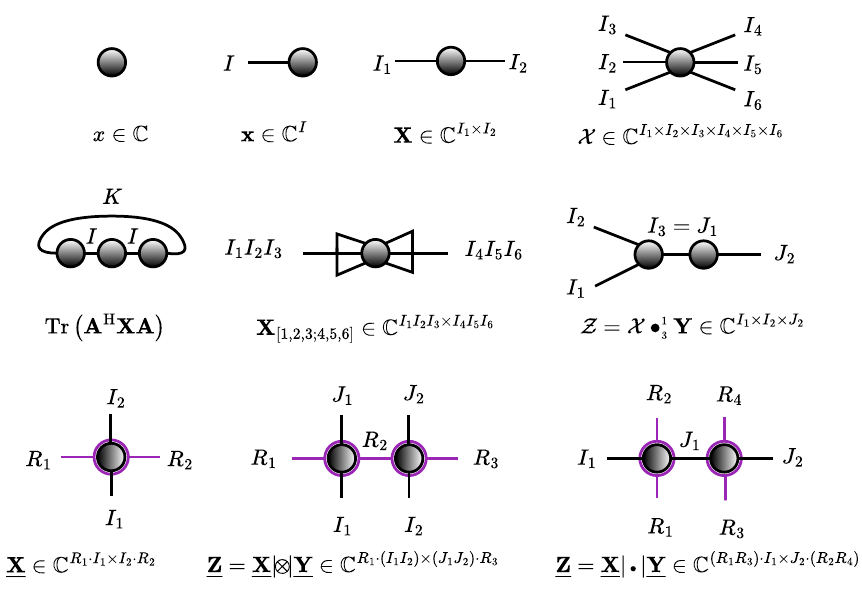}
	\caption{Basic tensor network diagrams. In the bottom row, double-ring nodes represent block matrices with purple block-layout edges and black inner-matrix edges; block matrices are linked by a purple edge in the strong Kronecker product and a black edge in the C-product.}
	\label{fignotation}
\vspace{-5mm}	
\end{figure}

We use lowercase, bold lowercase, bold uppercase, bold uppercase with an underline, and calligraphic letters to denote scalars, vectors, matrices, block matrices, and tensors, respectively; that is, $x, \vec{x}, \mat{x}, \ubar{\mat{x}} \text{ and } \ten{x}$. The trace of a matrix $\mat{x}$ is denoted by $\operatorname{Tr}(\mat{x})$. The Frobenius norm and the nuclear norm (trace norm) are denoted by \(\|\cdot\|_{\mathrm{F}}\) and \(\|\cdot\|_\star\), respectively.  The Kronecker delta is defined by $\delta_{ij}\!=\!1$ for $i\!=\!j$ and $0$ otherwise.  A state vector is denoted by \(\vec{\psi}\), a density matrix by \(\rho\), and a $d$-dimensional Hilbert space by \(\mathcal{H}_d\). The space of density matrices is denoted by $\mathcal{S}$, where $\mathcal{S} = \{ \rho \in \mathbb{C}^{D \times D} \mid \rho = \rho^{\mathrm{H}}, ~ \rho \succeq 0,~ \operatorname{Tr}(\rho) = 1\}$.  The outer product $\ten{Z} = \mat{X} \op \mat{Y} \in  \mathbb{C}^{I_{1} \times I_{2} \times J_{1} \times J_{2}}$ of two matrices $\mat{X} \in \C^{I_1 \times I_2}$ and $\mat{Y} \in \C^{J_1 \times J_2}$ satisfies $z_{ijkl} = x_{ij} y_{kl}$ and the Kronecker product is denoted by  $\mat{Z} = \mat{X} \otimes \mat{Y} \in \mathbb{C}^{I_{1}J_{1} \times I_{2}J_{2}}$. The $k$-fold Kronecker product of $\mat{X}$ is denoted by  $\mat{X}^{\otimes k} = \mat{X} \otimes \cdots \otimes \mat{X} \in \mathbb{C}^{I_1^k \times I_2^k}$.  The {\em order} of a tensor $\ten{x} \in \C^{I_{1} \times I_{2} \times \cdots \times I_{N}}$ is the number of its {\em modes,} which corresponds to the number of free edges in the tensor network diagram, as shown in \cref{fignotation}. A specific entry of an $N$th-order tensor is denoted by $x_{i_{1} i_{2} \cdots i_{N}}= \ten{x}(i_{1}, i_{2}, \ldots, i_{N})$. The $n$th matrix unfolding of an $N$th-order tensor is the matrix $\mat{x}_{[1, \ldots, n ; n\!+\!1, \ldots, N]} \in \C^{I_1\cdots I_n \times I_{n\!+\!1} \cdots I_N}$ obtained by reshaping the tensor such that the first $n$ modes form the row index and the remaining modes form the column index. The ordering of the indices determines how the tensor entries are stacked in the resulting matrix. For a tensor $\mathcal{X}\in\mathbb{C}^{R_1\times I_1\times J_1\times R_2}$ and its equivalent block-matrix representation $\ubar{\mathbf{X}}=[\mathbf{X}_{r_1r_2}] \in \mathbb{C}^{R_1  \cdot I_1 \times J_1  \cdot R_2}$, with blocks $\mathbf{X}_{r_1r_2}\in\mathbb{C}^{I_1\times J_1}$, we use $\mathbf{X}_{[1,2;3,4]}\in\mathbb{C}^{R_1I_1\times J_1R_2}$ for the corresponding matrix unfolding. The use of $``\cdot"$ in the dimension expressions indicates a block-matrix structure, with external indices defining the block layout and internal dimensions specifying the block size.  For a block matrix \(\ubar{\mat{X}} \in \mathbb{C}^{R_{1}\cdot I_1 \times J_1\cdot R_2}\), define the {\em left} and {\em right unfoldings} as \( \matL{X}=\mat{X}_{[1,2,3;4]}\in\mathbb{C}^{R_{1}I_1J_1\times R_2} \) and \( \matR{X}=\mat{X}_{[1;2,3,4]}\in\mathbb{C}^{R_{1}\times I_1J_1R_2}. \) The block matrix is called {\em left-} or {\em right-orthogonal} if \( \matL{X}^{\mathrm H}\matL{X}=\mat{I}_{R_2} \) or \( \matR{X} \matR{X}^{\mathrm H} =\mat{I}_{R_{1}}, \) respectively.  The blockwise conjugate transpose of $\ubar{\mat{X}} \in \mathbb{C}^{R_1\cdot I_1 \times J_1 \cdot R_2}$ is denoted by $\ubar{\mat{X}}^{\mathrm{H}} \in \mathbb{C}^{R_1 \cdot J_1 \times I_1 \cdot R_2}$ (see \cite{lee2018tenops} for details). $\ttm{X}$ denotes the matrix TT (TTM) decomposition of a matrix $\mat{X} \in \mathbb{C}^{\prod_{n\!=\!1}^N  I_n \times \prod_{n\!=\!1}^N J_n}$; see \eqref{eqn:ttm} for a detailed definition. $\ubar{\mat{X}}^{(<n)}$ and $\ubar{\mat{X}}^{(>n)}$ denote the left and right-interface matrices, respectively, and $\mat{X}_{\neq n}$ the frame matrix; see \Cref{subsce:ttreps}.

\begin{definition}[Contraction]
	A contraction of two tensors $\ten{x} \in \C^{I_1 \times I_2 \times \cdots \times I_N}$ and $\ten{y} \in \C^{J_1 \times J_2 \times \cdots \times J_M},$ with common modes $I_n\!=\!J_m\!=\!K,$ yields an $(N\!+\!M\!-\!2)$-order tensor $\ten{z}=\ten{x} \bullet_{\scriptscriptstyle{n}}^{\scriptscriptstyle{m}} \ten{y},$ the entries of which are given by: 
	\begin{equation*}
	z_{i_1 \ldots i_{n\!-\!1} i_{n\!+\!1} \ldots i_{N} j_1 \ldots j_{m\!-\!1} j_{m\!+\!1} \ldots j_{M}}=\sum_{k=1}^{K} x_{i_1 \ldots i_{n\!-\!1} k i_{n\!+\!1} \ldots i_{N}} y_{j_1 \ldots j_{m\!-\!1} k  j_{m\!+\!1} \ldots j_{M}}. 
	\end{equation*}
This is a fundamental operation in tensor networks and can be viewed as a multilinear extension of matrix–matrix multiplication.
\end{definition}
\begin{definition}[Strong Kronecker product]\label{def:skpproduct}
	Let $\ubar{\mat{X}} = [\mat{X}_{r_1 r_2}] \in \mathbb{C}^{R_1  \cdot I_1 \times J_1  \cdot R_2}$ and $\ubar{\mat{Y}} = [\mat{Y}_{r_2 r_3}] \in \mathbb{C}^{R_2  \cdot I_2 \times  J_2  \cdot R_3}$ be block matrices with $\mat{X}_{r_1 r_2} \in \mathbb{C}^{I_1 \times J_1}$ and $\mat{Y}_{r_2 r_3} \in \mathbb{C}^{I_2 \times J_2}$. The strong Kronecker product is defined by $\ubar{\mat{Z}} = \ubar{\mat{X}} \sktensor \ubar{\mat{Y}} \in \mathbb{C}^{R_1  \cdot (I_1 I_2) \times  (J_1 J_2)  \cdot R_3}$ with blocks \cite{cichocki2016tensor, lee2018tenops} 
	\begin{equation*}
	\mat{Z}_{r_1 r_3} = \sum_{r_2=1}^{R_2} \mat{X}_{r_1 r_2} \otimes \mat{Y}_{r_2 r_3}. 
	\end{equation*}
	This operation is analogous to block matrix multiplication, but with Kronecker products instead of matrix products. 
\end{definition}

\begin{definition}[Core (C) product]\label{def:cproduct}  Let $\ubar{\mat{X}} = [\mat{X}_{r_1 r_2}] \in \mathbb{C}^{R_1  \cdot I_1 \times J_1  \cdot R_2}$ and $\ubar{\mat{Y}} = [\mat{Y}_{r_3 r_4}] \in \mathbb{C}^{R_3  \cdot J_1 \times  J_2  \cdot R_4}$ be block matrices with $\mat{X}_{r_1 r_2} \in \mathbb{C}^{I_1 \times J_1}$ and $\mat{Y}_{r_3 r_4} \in \mathbb{C}^{J_1 \times J_2}$. The C-product is defined by $\ubar{\mat{Z}} = \ubar{\mat{X}} \sbullet \ubar{\mat{Y}} \in \mathbb{C}^{(R_1  R_3)  \cdot I_1  \times  J_2  \cdot (R_2 R_4)}$ with blocks \cite{cichocki2016tensor, lee2018tenops}
	\begin{equation*}
	\mat{Z}_{(r_1 r_3 ) (r_2 r_4)} = \mat{X}_{r_1 r_2} \mat{Y}_{r_3 r_4}. 
	\end{equation*}
This operation follows a standard Kronecker layout across the outer block indices, while combining the inner sub-blocks via standard matrix multiplication.
\end{definition}

For ease of understanding, we use tensor network diagrams to visualize tensors and their operations; \cref{fignotation} shows a few basic examples.

\unskip
\subsection{Quantum states, measurements  and  tomography} 

This section reviews quantum mechanical concepts like state representation, measurement operators, and existing low-rank QST approaches that motivate the Block-TT approach presented in  \Cref{sec:bttqst}.

\subsubsection*{State representation}
{A \emph{qubit} (quantum bit) is the fundamental unit of quantum information. Unlike a classical bit, which takes a definite value \(0\) or \(1\), a qubit can exist in a superposition of both. Any pure qubit state can be represented as a unit vector in a two-dimensional Hilbert space as \( |\vec{\psi}\rangle = a|\vec{0}\rangle + b|\vec{1}\rangle \), where \( |\bm{0}\rangle \) and \( |\bm{1}\rangle \)  form an orthonormal basis (written in Dirac notation) representing two distinct physical states of the qubit. The coefficients \( a,b \in \mathbb{C} \) are probability amplitudes that specify the contribution of $\vec{\psi}$ along the basis directions \(|\bm{0}\rangle \) and \( |\bm{1}\rangle \).  When a measurement is performed in this basis, the state collapses to either \( |\bm{0}\rangle \) or \( |\bm{1}\rangle \). However, in a random experiment, the probability of obtaining \( |\bm{0}\rangle \) is $|a|^2$, and the probability of obtaining \( |\bm{1}\rangle \) is  $|b|^2$, according to the Born rule.  The normalization condition \( |a|^2 \!+\! |b|^2 \!=\! 1 \) ensures that the total probability sums to one.  Equivalently, the state vector can be represented as, \( \vec{\psi} = \begin{bmatrix} a \\ b \end{bmatrix} \in \ten{H}_2 \!\coloneq\!  \mathbb{C}^2 \).  A \emph{qudit} is a $d$-dimensional extension of a qubit, with pure states given by unit vectors in \(\mathbb{C}^d\).  A pure state of an $N$-qudit system is described by unit vectors of a Hilbert space formed by the tensor product of constituent Hilbert spaces.  For more details, we refer to \cite{paris2004qse,nielsen2010quantum, paris2012modern, thew2002qudit, qin2026QSTreview}. 

Thus far, we have only discussed pure states; a quantum system may also exist in a probabilistic mixture of pure states, in which the system occupies one of several \(\vec{\psi}_k\) with associated probabilities \(\alpha_k \). That is, the system is described by the statistical ensemble $\{ \alpha_k, \vec{\psi}_k\}$.  A more general description of the (mixed) state is given by a statistical operator (density matrix)  \( \rho \!=\! \sum_k \alpha_k \vec{\psi}_k \vec{\psi}_k^{\ast} \), where \( \alpha_k \ge 0 \), \( \sum_k \alpha_k \!=\! 1 \), and \(\vec{\psi}_k \) are the (pure) state vectors.  Following standard terminology in quantum information theory, we use ``state" and ``density matrix" in the rest of the paper interchangeably.

\subsubsection{Quantum measurements} According to the basic postulates of quantum mechanics, observable (measurable) quantities of a quantum system are represented by Hermitian operators. Any such operator  \(\mat{M}\!=\!\mat{M}^{\mathrm H}\) admits a spectral decomposition \(\mat{M}=\sum_k \lambda_k \mat{P}_{k}\), where the \(\lambda_k\) are real eigenvalues representing the possible outcomes of a measurement, and \(\mat{P}_{k}\) are orthogonal projectors onto the corresponding eigenspaces, satisfying \( \mat{P}_{k}\mat{P}_{l} = \delta_{kl}\mat{P}_k \) (orthogonality) and \( \sum_k \mat{P}_{k} = \mat{I}\) (completeness) \cite{paris2004qse,nielsen2010quantum, paris2012modern}. The probability $p_{\lambda_k}$ of obtaining outcome \(\lambda_k\) is given by $\operatorname{Tr}\left( \rho \mat{P}_{k} \right)$, and the overall expectation value is $\operatorname{Tr}\left( \rho \mat{M} \right)$.  The  completeness condition ensures that the measurement accounts for all possible outcomes and produces a properly normalized probability distribution\footnote{That is $\sum_k p_{\lambda_k} =  \sum_k \mathrm{Tr}(\rho \mat{P}_{k} ) =\operatorname{Tr}\!\left(\rho \sum_k \mat{P}_{k} \right) = \mathrm{Tr}(\rho \mat{I}) = \mathrm{Tr}(\rho) = 1.$}. A measurement defined by orthogonal projectors $\{\mat{P}_k\}$ satisfying completeness, and inducing a probability measure on outcomes, is commonly called a projective-valued measure (PVM) \cite{paris2004qse, nielsen2010quantum}.

Notice that, in a projective measurement framework, the number of outcomes is limited by the Hilbert space dimension, as there can be at most that many mutually orthogonal projectors. This limitation is addressed by a more general measurement framework, known as the positive operator-valued measure (POVM), which relaxes the orthogonality and projection requirements of PVM elements, replacing them with the weaker conditions of positivity and completeness, while still encoding the valid and meaningful probabilistic nature of the measurement outcomes \cite{nielsen2010quantum, paris2012modern}. That is, a POVM is a collection of $K$ measurement operators $\{\mat{E}_k\}_{k=1}^K$ satisfying $\mat{E}_k \succcurlyeq 0$ (positivity) and $\sum_{k\!=\!1}^K \mat{E}_k = \mat{I}$ (completeness).  As a consequence, a POVM can have more elements than the dimension of the Hilbert space they act in.  Each element $\mat{E}_k$  in the POVM is associated with a possible outcome of a quantum measurement. The probability $p_{k}$  of getting the outcome with label $k$ is given by $\operatorname{Tr}\left( \rho \mat{E}_{k} \right)$, so a single-shot measurement yields label \(k\) at random with this probability. Consequently, estimating all \(\{p_k\}\) requires multiple shots on identically prepared copies of the state, from which the probabilities are inferred via empirical frequencies \cite{paris2004qse, qin2026QSTreview}. 

A POVM is {\em informationally complete} if its elements contain enough information to uniquely infer the state from the outcomes. Otherwise, one can combine multiple POVMs—say \(L\) of them—so that the union of the resulting \(M  \!=\! KL\) elements is informationally complete.  A notable difference between POVMs and PVMs lies in their ability to be informationally complete. For a generic mixed state, the density matrix has a number of independent parameters that scales quadratically with the Hilbert space dimension. A single PVM, however, yields at most a number of independent outcomes equal to that dimension and is therefore never informationally complete. In contrast, a single POVM is not subject to this restriction; a single POVM can therefore be informationally complete.  See \cite{paris2004qse,nielsen2010quantum, paris2012modern, qin2026QSTreview} for more details. \par

	The \emph{Pauli spin matrices} for a single qubit are denoted by $\mat{I}_2, \bm{\sigma}_x:=\begin{bmatrix} 0 & 1 \\ 1 & 0 \end{bmatrix}, \bm{\sigma}_y\coloneq \begin{bmatrix} 0 & -i \\ i & 0 \end{bmatrix}, \bm{\sigma}_z \coloneq \begin{bmatrix} 1 & 0 \\ 0 & -1 \end{bmatrix}$.} They form an informationally complete set for a single-qubit  system. Notice that this set of matrices is not a POVM itself, but measuring 
	each of these observables is described by a two‑outcome POVM of the form  \(\{\frac{\mat{I}_2 \!+\! \bm{\sigma}_n}{2},  \frac{\mat{I}_2 \!-\! \bm{\sigma}_n}{2} \}, 
	\) where \(\bm{\sigma}_n \in \{\mat{I}_2, \bm{\sigma}_x, \bm{\sigma}_y, \bm{\sigma}_z\}\).  Pauli matrices are  commonly used as measurement observables in 
	single‑qubit systems. For an $N$-qubit system,  local\footnote{That is, acting independently on individual qubits.} measurement observables can be formed as 
	Kronecker products of single‑qubit Pauli matrices,
	\(
	\mat{E}_m \!:=\! \otimes_{n\!=\!1}^{N} \bm{\sigma}_{n},
	\)  where $ \bm{\sigma}_{n}$  denotes the Pauli operator acting on qubit $n$. \par
	
	Another commonly used local (i.e., single-qubit) measurement is the symmetric informationally complete (SIC) POVM \cite{renes2004SICPOVM}, given by
	\begin{equation*}
	\left\{
	\begin{bmatrix}
\frac{1}{2} & 0 \\
0 & 0
\end{bmatrix},
\;
\begin{bmatrix}
\frac{1}{6} & \frac{\sqrt{2}}{6} \\
\frac{\sqrt{2}}{6} & \frac{1}{6}
\end{bmatrix},
\;
\begin{bmatrix}
\frac{1}{6} & \frac{\sqrt{2}}{6} e^{-i\frac{2\pi}{3}} \\
\frac{\sqrt{2}}{6} e^{i\frac{2\pi}{3}} & \frac{1}{6}
\end{bmatrix},
\;
\begin{bmatrix}
\frac{1}{6} & \frac{\sqrt{2}}{6} e^{-i\frac{4\pi}{3}} \\
\frac{\sqrt{2}}{6} e^{i\frac{4\pi}{3}} & \frac{1}{6}
\end{bmatrix}
\right\}.
\end{equation*}

\subsubsection*{Low-rank QST}\label{subsec:lrqst}  QST can be formulated as an inverse problem. Consider an $N$-qudit system defined on the Hilbert space \(\ten{H}_{D} \!\coloneq\! (\mathbb{C}^d)^{\otimes N},\) with total dimension $D \!=\! d^N$. The state of the system is described by a density matrix \(\rho \in \mathbb{C}^{D \times D},\) belonging to $\mathcal{S}$, the set of Hermitian, PSD matrices with unit trace. In the special case of qubits, we have $d \!=\! 2$. Suppose we measure an informationally complete set of POVM elements, obtaining (possibly noisy) measurement statistics $\{\mathbf{E}_m, y_m\}_{m\!=\!1}^{M}$. The goal of QST is to reconstruct a density matrix $\hat{\rho}$ consistent with the observations, i.e., $y_m \approx \operatorname{Tr}(\hat{\rho} \mat{E}_m), \text{ for } m\!=\!1, 2, \ldots, M$. This recovery problem is challenging because (1) $\hat{\rho}$ must satisfy the physicality constraints, i.e., $\hat{\rho} \in \mathcal{S}$, and (2) the number of measurement settings required to ensure unique recovery of a generic, unstructured density matrix scales as $\mathcal{O}(D^2)$ \cite{wang2013qstvmc}. However, if the underlying density matrix is low-rank, it has been shown that significantly fewer measurements suffice to estimate it. Under Pauli measurements, the rank-$R$ matrix has been shown to be uniquely estimated using $\mathcal{O}(R D \log^2 D)$ measurement settings, with high probability \cite{gross2010csqst, liu2011universal, liu2012csqst}. Under structured measurements, $\mathcal{O}(R D)$ settings may suffice \cite{baldwin2016strictly, tariq2024efficient, sofi2026tomography}. 
The literature outlines the following two approaches to employing the low-rank constraints.

\paragraph{Rank minimization}{In this method, the low-rank constraint is enforced implicitly by formulating a rank-minimization objective (with nuclear norm relaxation of the rank), under measurements and physicality constraints. Mathematically \cite{gross2010csqst},
	\begin{equation}
		\label{eqn:sdpQST}
		\min_{\hat{\rho} \in \mathbb{C}^{D \times D}} \|\hat{\rho}\|_{\star} \quad \text{s.t.} \quad \|\vec{y} - \ten{M}(\hat{\rho})\|_2 \leq \epsilon, \text{ and } \hat{\rho} \succeq 0, 
	\end{equation}
where $\vec{y} \in \mathbb{R}^M$ are the measurement outcomes, and the measurement (sensing) map $\mathcal{M}$ is $(\mathcal{M}(\rho))_m = \operatorname{Tr}(\mat{E}_m \rho)$. To ensure $\hat{\rho}\in\mathcal{S}$, one generally normalizes the solution as $\hat{\rho}:=\hat{\rho}/\operatorname{Tr}(\hat{\rho})$; for noiseless, trace-preserving measurements, the unit-trace condition is automatically satisfied under the recovery conditions \cite{kalev2015quantum}.  The nuclear norm relaxation is the tightest convex surrogate of the rank function; directly minimizing the rank leads to a non-convex and NP-hard problem \cite{boyd1997sdp, fazel2004rank}. This relaxation enables an approximate yet computationally more efficient solution. The problem \eqref{eqn:sdpQST} can be formulated as a semidefinite program (SDP) \cite{fazel2002matrix, fazel2004rank}, and when $N$ is small, the SDP can be readily solved using software such as SDPT3 or  CVXPY \cite{cvxpy}. For mid- to large-scale systems, optimizing over a full $(D \times D)$ density matrix becomes infeasible.}

\paragraph{Error minimization}{ This approach explicitly promotes low-rank structure by parametrizing the density matrix using a fixed, low-rank Burer--Monteiro  (Cholesky-like) factorization \cite{burer2003lrsdp}. While this parameterization renders the problem non-convex, it significantly improves computational efficiency by restricting optimization to low-rank factors. Mathematically \cite{burer2003lrsdp}, 
	\begin{equation}
		\label{eqn:BMQST}
		\min_{\mat{A} \in \mathbb{C}^{D \times R}} \| \vec{y}  - \ten{M}\left(\hat{\rho}\right) \|_2^2 ~ \text { with } \hat{\rho} = \mat{A}\mat{A}^{\mathrm{H}},\text { s.t. }  \operatorname{Tr}(\hat{\rho})=1. 
	\end{equation}
Here, the flexibility to choose a differentiable objective function allows for the use of gradient-based optimization approaches (e.g., first- and/or second-order methods), yielding an efficient approximate solution to QST. Moreover, the Hermitian and PSD constraints are automatically satisfied by construction. In general, the optimization problems arising from this parameterization may admit local solutions\footnote{Without any additional constraints beyond the PSD constraint, one can decompose $\rho$ as $\rho  = \mat{\hat{A}} \mat{\hat{A}}^{\mathrm{H}},$ where $\mat{\hat{A} = AL}$ for any unitary $\mat{L} \in \C^{R \times R}$ such that $\mat{LL}^{\mathrm{H}} = \mat{I}$.}. However, with mild regularity conditions and suitable initialization, this approach becomes effective for many applications, including QST \cite{bhojanapalli2016lrsdp, burer2003lrsdp, kalev2015quantum, kyrillidis2018provable, kim2023mifgd}. In practice, the unit-trace requirement is often relaxed to the convex constraint \( \operatorname{Tr}(\hat{\rho}) \leq 1 \;\Longleftrightarrow\;  \|\mat{A}\|_{F}^2 \leq 1 , \) to further improve scalability \cite{bhojanapalli2016lrsdp, kyrillidis2018provable}.  While the parameters of $\mat{A}$ remain exponential in $N$, our approach generalizes the Burer--Monteiro factorization to the Block-TT format, breaking this exponential dependence. \par
	
\subsection{Tensor train decomposition} \label{subsce:ttreps}

A TT decomposition of an $N$th-order tensor \(\ten{x} \in \C^{I_1 \times I_2 \times \cdots \times I_N}\) corresponds to a contraction of a sequence of third-order tensors (TT-cores) $\ten{x}^{(n)} \in \C^{R_{n\!-\!1} \times I_n  \times R_{n}} (1\!\leq \!n \!\leq \!N)$ with $R_{0}\!=\! R_{N}\!=\!1$ such that each entry of $\ten{x}$ can be expressed as the sequence of matrix products  \cite{verstraete2007MPS, oseledets2010tensortrain}:
	\begin{equation}
		\label{eqn:tt}
		x_{i_{1} i_{2} \cdots i_{N\!-\!1} i_{N}} = \mat{x}^{(1)}_{: i_{1} :} \mat{x}^{(2)}_{: i_{2} :}  \cdots \mat{x}^{(N\!-\!1)}_{: i_{N\!-\!1} :} \mat{x}^{(N)}_{: i_{N} :},
	\end{equation}
where the matrix $\mat{x}^{(n)}_{: i :} \in \C^{R_{n\!-\!1} \times R_{n}}$ is the $i$th mode-2 slice of the TT-core $\ten{x}^{(n)}.$ The tuple of minimal integers $(R_0, \ldots, R_{N})$ for which equality in~\Cref{eqn:tt} holds is the TT-rank of $\ten{x}$.\par
	
A dense tensor $\ten{x}$  can be decomposed into a TT format via the TT-SVD algorithm, which successively reshapes the tensor into unfolding matrices with one isolated mode and computes truncated SVDs from left to right. At each step, the left singular vectors form a TT-core, while the remaining factor is reshaped and propagated to the subsequent decomposition step \cite{oseledets2010tensortrain}. Due to its sequential nature, TT-SVD is difficult to parallelize; alternatively, Parallel-TTSVD computes truncated SVDs of all $n$th unfoldings independently and then constructs the TT-cores from them \cite{shi2023paralleltt}.
	
\subsubsection*{TT representations for vectors and matrices} \label{subsubsec:ttm}
Large-scale  vectors and matrices can be represented in TT format after tensorization. Specifically, a vector $\vec{x}\in\mathbb{C}^{I}$ is reshaped into a tensor $\ten{x}\in\mathbb{C}^{I_1\times\cdots\times I_N}$, while a matrix $\mat{X}\in\mathbb{C}^{I\times J}$ is reshaped into a tensor $\ten{X}\in\mathbb{C}^{(I_1J_1)\times\cdots\times(I_NJ_N)}$, where $I\!=\!\prod_{n\!=\!1}^{N} I_n$ and $J\!=\!\prod_{n\!=\!1}^{N} J_n$. The corresponding TT-cores are third-order tensors  $\ten{X}^{(n)}\in\mathbb{C}^{R_{n\!-\!1}\times I_n\times R_n}$ for vectors and fourth-order tensors $\ten{X}^{(n)}\in\mathbb{C}^{R_{n\!-\!1}\times I_n\times J_n\times R_n}$ for matrices, obtained by splitting the ``long mode"  $I_n J_n$  into $I_n\!\times \!J_n$ \cite{oseledets2010tensortrain, oseledets2010ttm}. In this paper, we adopt an equivalent TT representation using the strong Kronecker product, together with the C-product to describe tensor network contractions. This notation provides a more intuitive framework for visualizing operations on tensors obtained by tensorizing vectors and matrices.

In the matrix case, a matrix TT (TTM \cite{oseledets2010ttm}), mathematically equivalent to the MPO, can also be represented as a sequential strong Kronecker product of \(N\) block matrices,
\begin{equation}
	\label{eqn:ttm}
	\ttm{X} = \ubar{\mat{X}}^{(1)} \sktensor \ubar{\mat{X}}^{(2)} \sktensor \cdots \sktensor \ubar{\mat{X}}^{(N)} \in \C^{ \prod_{n\!=\!1}^N  I_n \times \prod_{n\!=\!1}^N J_n},
\end{equation}
where each core \(\ubar{\mat{X}}^{(n)}\in\mathbb{C}^{R_{n-1} \cdot I_n\times J_n  \cdot R_n}\) consists of blocks $\mat{X}^{(n)}_{r_{n\!-\!1} : : r_{n}} \in \C^{I_n \times J_n}$:
\begin{equation*}
\ubar{\mat{X}}^{(n)}
=
\begin{bmatrix}
	\mat{X}^{(n)}_{1 : : 1} & \cdots & \mat{X}^{(n)}_{1 : : R_n} \\
	\vdots & \ddots & \vdots \\
	\mat{X}^{(n)}_{R_{n\!-\!1} : : 1} & \cdots & \mat{X}^{(n)}_{R_{n\!-\!1} : : R_n}
\end{bmatrix}.
\end{equation*}
The corresponding tensor entries satisfy\footnote{The strong Kronecker product representation formally yields a matrix; here, entries are understood via the associated tensorized representation.}
\begin{equation}
	\label{eqn:ttmentry}
x_{i_1 j_1 i_2 j_2\ldots i_N  j_N} = \mat{X}^{(1)}_{:i_1j_1:} \mat{X}^{(2)}_{:i_2j_2:} \cdots \mat{X}^{(N)}_{:i_Nj_N:}.
\end{equation}
Note that \( \ubar{\mat{X}}^{(n)} \in \mathbb{C}^{R_{n\!-\!1}  \cdot I_n \times  J_n  \cdot R_{n} } \) is referred to as a \emph{core}, while its corresponding tensor in the TT decomposition \( \ten{X}^{(n)} \in \mathbb{C}^{R_{n\!-\!1} \times I_n \times  J_n \times R_{n} } \) is referred to as a \emph{TT-core}. The TT-rank \( (R_0, \ldots, R_N) \) corresponds to the number of blocks in \( \ubar{\mat{X}}^{(n)} \), with $R_{0}\!=\! R_{N}\!=\!1$.  Beyond its standard view as a tuple of unfolding matrix ranks quantifying row and column independence, the TT-rank explicitly captures the internal structural complexity of the strong Kronecker product representation.

 In the vector case \(J\!=\!1\) (and hence \(J_n\!=\!1,\ \forall n\)), each block of \(\ubar{\mat{X}}^{(n)}\) reduces to a vector of size \(I_n\), yielding the standard TT/MPS representation.
 
A Block$\!-\!n\!-\!$TT is a TTM where all cores consist of vector-valued blocks except for a single matrix-valued block core at position $n$ ($J_i\!=\!1$ for $i\!\neq\! n$, $J_n\!=\!K$) \cite{pivzorn2012variational, dolgov2014computation, kressner2014btt, lee2015btt}. As shown in \cref{figbtt}, it is equivalent to a collection of TT vectors  differing only in their $n$th core.

\begin{figure}[htb]
	\centering
     \includegraphics[width=0.75\linewidth]{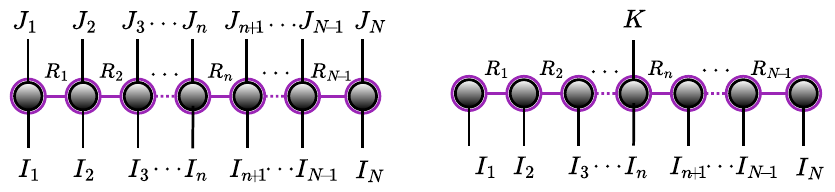}
	\caption{(Left)~TTM representation of a matrix $\mat{X} \in \C^{I_1 I_2 \cdots I_N \times J_1 J_2 \cdots J_N}$. (Right)~Block\(\!-\!n\!-\!\)TT representation of a tall matrix $\mat{X} \in \C^{I_1 I_2 \cdots I_N \times K}$, or equivalently, as a collection of $K$ TT vectors that share all cores, except for the $n$th core.  }
\label{figbtt}
\end{figure}

A TTM can also be represented using the shorthand notation
\begin{equation}
	\label{eqn:ttmshort}
	\ttm{X} = \ubar{\mat{X}}^{(<n)} \sktensor \ubar{\mat{X}}^{(n)} \sktensor \ubar{\mat{X}}^{(>n)} = \ubar{\mat{X}}^{(<n)} \sktensor \ubar{\mat{X}}^{(n, n\!+\!1)} \sktensor \ubar{\mat{X}}^{(>n\!+\!1)}, 
\end{equation}
where the left-interface  $\ubar{\mat{X}}^{(<n)} = \ubar{\mat{X}}^{(1)} \sktensor \cdots \sktensor \ubar{\mat{X}}^{(n\!-\!1)} \in \C^{R_0  \cdot \left(\prod_{i\!=\!1}^{(n\!-\!1)} I_i\right) \times \left(\prod_{i\!=\!1}^{(n\!-\!1)} J_i\right)  \cdot R_{n\!-\!1}},$  the right-interface $\ubar{\mat{X}}^{(>n)} = \ubar{\mat{X}}^{(n\!+\!1)} \sktensor \cdots \sktensor \ubar{\mat{X}}^{(N)} \in \C^{R_{n}  \cdot \left(\prod_{i\!=\!n\!+\!1}^{N} I_i\right) \times \left(\prod_{i\!=\!n\!+\!1}^{N} J_i\right)  \cdot R_N}$ and the merged core  $\ubar{\mat{X}}^{(n, n\!+\!1)} = \ubar{\mat{X}}^{(n)} \sktensor  \ubar{\mat{X}}^{(n\!+\!1)} \in \mathbb{C}^{R_{n-1} \cdot (I_n I_{n\!+\!1}) \times (J_n J_{n\!+\!1}) \cdot R_{n\!+\!1}}$. The merged core is commonly referred to as a ``supercore" or a ``two-site tensor" in the DMRG literature \cite{white1993density, schollwock2005dmrg, khoromskij2010dmrgqtt, oseledets2011dmrg,  holtz2012alternating, kressner2014btt, lee2015btt}.\par 

If two TTMs $\ttm{X} = \ubar{\mat{X}}^{(1)} \sktensor  \cdots \sktensor \ubar{\mat{X}}^{(N)}$ and $\ttm{Y} = \ubar{\mat{Y}}^{(1)} \sktensor  \cdots \sktensor \ubar{\mat{Y}}^{(N)}$  have the same mode sizes, their linear combination $\ttm{Z} = \alpha \ttm{X} + \beta \ttm{Y}$, can be expressed as 
\begin{equation*}
\ttm{Z}
=
\bigl[ \ubar{\mat{X}}^{(1)}  \  \ubar{\mat{Y}}^{(1)} \bigr]
\sktensor
\begin{bmatrix}
\ubar{\mat{X}}^{(2)}  & \bm{0} \\
\bm{0} &  \ubar{\mat{Y}}^{(2)} 
\end{bmatrix}
\sktensor \cdots \sktensor
\begin{bmatrix}
 \ubar{\mat{X}}^{(N\!-\!1)}  & \bm{0} \\
\bm{0} & \ubar{\mat{Y}}^{(N\!-\!1)}
\end{bmatrix}
\sktensor
\begin{bmatrix}
\alpha \ubar{\mat{X}}^{(N)} \\
\beta \ubar{\mat{Y}}^{(N)}
\end{bmatrix}. 
\end{equation*}
The multiplication $\ttm{Z} = \ttm{X}\ttm{Y}$, with $\ttm{X}$ and $\ttm{Y}$ having compatible mode sizes, can be compactly written as 
\begin{equation*}
\ttm{Z} = \left(\ubar{\mat{X}}^{(1)} \sbullet \ubar{\mat{Y}}^{(1)} \right) \sktensor \cdots  \sktensor \left(\ubar{\mat{X}}^{(N)} \sbullet \ubar{\mat{Y}}^{(N)} \right). \vspace{-2mm}
\end{equation*}
Notice that TTM addition concatenates cores block-diagonally (direct sum), upper-bounding the ranks by the sum of the corresponding core ranks, whereas TTM multiplication combines cores via C-products, upper-bounding the ranks by their product. In practice, one performs the TT-rounding operation \cite{oseledets2010tensortrain} to keep the ranks minimal.

\begin{definition}[Frame equation for Block-TT]\label{def:frame}  
A TT(M) representation is multilinear with respect to its cores. Consequently, the Block\(\!-\!n\!-\!\)TT representation ($J_i\!=\!1$ for $i\!\neq\! n$, $J_n\!=\!K$) is linear with respect to the $n$th core. The same property
holds for a merged two-site core.  It can be expressed
using the matrix frame equation, as defined in
\cite{kressner2014btt, dolgov2014computation, lee2015btt}:
\begin{equation}\label{eqn:framebtt}
	\ttm{X} = \mat{X}_{\neq n}\; \mat{X}^{(n)}_{[1,2,4;3]} = \mat{X}_{\neq n, n\!+\!1}\; \mat{X}^{(n, n\!+\!1)}_{[1,2,4;3]} \in \C^{I_1 I_2 \cdots I_N \times K}.  
	\end{equation}
The corresponding frame matrices are constructed from the left and right
interfaces as \(
	\mat{X}_{\neq n}
	= \matL{X}^{(<n)} \otimes \mat{I}_{I_n} \otimes \matR{X}^{(>n) \mathrm{H}}\in \C^{\left(\prod_{i\!=\!1}^{N} I_i  \right) \times \left(R_{n\!-\!1} I_n R_n \right) },
	\) and 
	\(
	\mat{X}_{\neq n, n\!+\!1}
	= \matL{X}^{(<n)} \otimes \mat{I}_{I_n}  \otimes \mat{I}_{I_{n\!+\!1}} \otimes \matR{X}^{(>n\!+\!1) \mathrm{H}} \in \C^{\left(\prod_{i\!=\!1}^{N} I_i J_i \right) \times \left(R_{n\!-\!1} I_n I_{n\!+\!1} R_{n\!+\!1} \right) }.
	\)  Here, $\mat{X}^{(n)}_{[1,2,4;3]}\in
	\mathbb{C}^{R_{n\!-\!1}I_nR_n\times K}$ and
	$\mat{X}^{(n,n\!+\!1)}_{[1,2,4;3]}\in
	\mathbb{C}^{R_{n\!-\!1}I_nI_{n\!+\!1}R_{n\!+\!1}\times K}$
	are the  matrix unfoldings of the core
	$\ubar{\mat{X}}^{(n)}$ and the merged two-site core
	$\ubar{\mat{X}}^{(n,n\!+\!1)}$, respectively.
\end{definition}

\begin{definition}[\(n\!-\!\)orthogonal form]\label{def:northo}
A TT(M) is in \(n\!-\!\)orthogonal form if all cores to the left of the $n$th core are left-orthogonal and all cores to its right are right-orthogonal. To obtain the \(n\!-\!\)orthogonality, one proceeds sequentially from the first (or last) core. At each step, a QR (or LQ) decomposition is applied to the left (or right) unfolding of the \(i\)th  core. The orthogonal factor is reshaped back into a core, while the triangular factor is absorbed into the next (or previous) core. This procedure is repeated until the \(n\!-\!\)orthogonal form is achieved \cite{oseledets2010tensortrain, schollwock2005dmrg, white2005density, holtz2012alternating, kressner2014btt, dolgov2014computation}. Notice that \(n\!-\!\)orthogonality ensures orthogonality of the left- and right-interface matrices, i.e., 
\begin{equation*}
\matL{X}^{(<n) \mathrm{H}}
\matL{X}^{(<n)} = \mat{I}_{R_{n\!-\!1}},
\quad  \quad
\matR{X}^{(>n)}
\matR{X}^{(>n) \mathrm{H}}
= \mat{I}_{R_{n}}. \vspace{-1.5mm}
\end{equation*}
Consequently, the frame matrices $\mat{X}_{\neq n}$ and $\mat{X}_{\neq n, n\!+\!1}$ are column-wise orthogonal. 
In the literature, $1$-orthogonal and $N$-orthogonal forms are also referred to as right-orthogonal and left-orthogonal forms, respectively.

\end{definition}

\section{Block-TT approach to low-rank QST} \label{sec:bttqst}
This section presents our main contribution. We use an efficient extension of the Burer–Monteiro factorization to tensors to circumvent the curse of dimensionality in QST. More specifically, we show in \Cref{subsec:bttden} that the density matrix can be parameterized as a contraction of two Block-TT networks, thereby enforcing Hermiticity and positivity by construction. This parameterization enables a DMRG-like optimization scheme for QST of low-rank mixed states, which is developed in \Cref{subsubsec:dmrgqst}.

\subsection{Block-TT for mixed low-rank states}\label{subsec:bttden}
We employ a representation that expresses a density matrix as a contraction of a Block$\!-\!n\!-\!$TT network with its Hermitian transpose, i.e., 
\begin{equation}
		\label{eqn:BTTrho}
\ttm{\rho} = \ttm{A} \ttm{A}^{\mathrm{H}} \in \mathbb{C}^{d^N \times d^N}\!, \quad \ttm{A} = \ubar{\mat{A}}^{(1)} \sktensor \ubar{\mat{A}}^{(2)} \sktensor \cdots \sktensor \ubar{\mat{A}}^{(N)} \in \mathbb{C}^{d^N \times K}, 
\end{equation}
or entry-wise, we can write: 
\begin{equation}
\label{eqn:BTTrho2}
\oldrho_{{\scriptscriptstyle \mathsf{TT}}\, i_1j_1\ldots i_N j_N} = \mat{\rho}^{(1)}_{:i_1j_1:} \, \mat{\rho}^{(2)}_{:i_2j_2:} \, \cdots \, \mat{\rho}^{(N)}_{:i_Nj_N:},
\quad
\ubar{\rho}^{(i)} = \ubar{\mat{A}}^{(i)} \sbullet \ubar{\mat{A}}^{(i)\mathrm{H}}. 
\end{equation}
Here, $\ttm{A} \in \mathbb{C}^{d^N \times K}$ bounds the matrix rank of $\ttm{\rho}$ by $K$. Cores  $\ubar{\mat{A}}^{(i)} \in \mathbb{C}^{R_{i\!-\!1} \cdot d \times R_i}$ are vector-valued block cores for $i \!\neq\! n$, while the $n$th core is a matrix-valued block core,
$\ubar{\mat{A}}^{(n)} \in \mathbb{C}^{R_{n\!-\!1} \cdot d \times K \cdot R_n}$.   Intuitively, this construction is analogous to the Burer–Monteiro factorization for matrices \cite{burer2003lrsdp}. A tensor network diagram of this model is shown in \cref{figbttden}.

\begin{figure}[htb]
	\centering
     \includegraphics[width=0.5\linewidth]{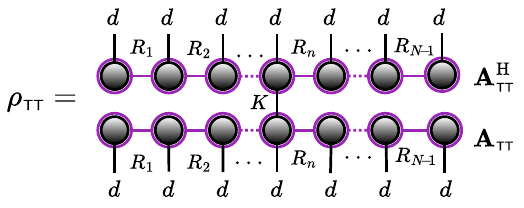}
	\caption{Burer--Monteiro parameterization of a density matrix in Block$\!-\!n\!-\!$TT format, i.e., $\ttm{\rho} = \ttm{A}\ttm{A}^{\mathrm{H}}$, where $\ttm{A}\in \mathbb{C}^{D \times K}, \, D\!=\!d^N$.}
\label{figbttden}
\end{figure}

\subsubsection*{Physicality constraints} \label{par:physicality}
As noted earlier, a Block-TT can be interpreted as a collection of TT vectors sharing all cores except the $n$th core (i.e., the core carrying the block index $K$). The corresponding frame equation can be written in the following form (cf.~\Cref{eqn:framebtt}):
\(
\ttm{A} = \mat{A}_{\neq n}\, \mat{A}^{(n)}_{[1,2,4;3]},
\)
where the reshaped $n$th core is
\(
\mat{A}^{(n)}_{[1,2,4;3]} \in \C^{ R_{n\!-\!1} d R_n \times K},
\)
and the frame matrix is given by
\(
\mat{A}_{\neq n}
=
\matL{A}^{(<n)}
\otimes
\mat{I}_{d}
\otimes
\matR{A}^{(>n)\mathrm{H}}
\in
\C^{D\times R_{n\!-\!1} d R_n}.
\) Assume we work under $n$-orthogonality (see ~\cref{def:northo}). Then the spectrum of $\ttm{\rho}$ satisfies 
\begin{equation} \label{eqn:phy}
\sigma\!\left(\ttm{A} \ttm{A}^{\mathrm{H}}\right)
=
\sigma\!\left(\mat{A}_{\neq n}\,\mat{A}^{(n)}_{[1,2,4;3]}  \mat{A}^{(n)\mathrm{H}}_{[1,2,4;3]}\, \mat{A}_{\neq n}^{\mathrm{H}} \right)
= \sigma\!\left(\mat{A}_{\neq n}\, \eff{\rho}{n}\,\mat{A}_{\neq n}^{\mathrm{H}} \right) = 
\sigma\!\left(\eff{\rho}{n} \right). 
\end{equation}
Therefore, the eigenvalues of the effective density matrix
\(\eff{\rho}{n} = \mat{A}^{(n)}_{[1,2,4;3]}  \mat{A}^{(n)\mathrm{H}}_{[1,2,4;3]} \in \C^{R_{n\!-\!1} d R_n \times R_{n\!-\!1} d R_n}\)
coincide with the nonzero eigenvalues of $\ttm{\rho}$. Consequently, the normalization constraints (unit-trace) also follow from local constraints, i.e., $\operatorname{Tr}\!\left(\ttm{\rho}\right)
= \operatorname{Tr}\!\left(\eff{\rho}{n}\right) = 1$. Notice that the effective density matrix has a Gram structure and therefore guarantees that $\ttm{\rho}$ is positive semidefinite by construction.} \par

We observe that, under the $n$-orthogonality constraints, the structural properties of the global density matrix
\(
\ttm{\rho} = \mat{A}_{\neq n}\,\eff{\rho}{n}\, \mat{A}_{\neq n}^{\mathrm{H}}
\)
are reflected in the effective density matrix. Owing to this renormalization-type (scale/transformation-invariant) structure, an SDP optimization problem posed on $\ttm{\rho}$ can be reduced to a local problem on $\eff{\rho}{n}$, which preserves the SDP structure provided that $n$-orthogonality holds and optimization constraints map linearly under frame projections. Note, however, that we do not solve the SDP for $\eff{\rho}{n}$ directly. Instead, we optimize over the low-rank factors $\mat{A}^{(n)}_{[1,2,4;3]}$, which results in a nonconvex optimization problem. After solving the local problem, the orthogonality center can be moved to the next (or previous) core, for instance, by shifting the core carrying the block index $K$ and applying a QR (or LQ) step, as in~\cref{def:northo}. See the \hyperref[par:rankadaptivity]{rank adaptation and block-index sweeping} in \Cref{subsubsec:dmrgqst} for details.

We emphasize that the Block-TT parameterization of the density matrix is efficient only when the mixedness is low, i.e., when the density matrix can be expressed as a sum of a few pure states. This parametrization allows for representing a mixed state with asymptotically the same number of parameters as a TT/MPS. That is, the number of parameters is equal to  $dR^2 (\log_dD-3) +  dR^2 K + 2dR $, assuming $R_n=R$.

\subsection{Compressed sensing approach}\label{subsec:csqst} 
We extend the error-minimization low-rank QST formulation \eqref{eqn:BMQST} by parameterizing the density matrix via Block-TT network contractions. This approach offers two main advantages: logarithmic parameter scaling in the state dimension and fast expectation-value evaluation via tensor-network contractions. The low-rank Burer--Monteiro QST is formulated as:
\begin{equation}
\label{eqn:bttqst}
\displaystyle \min_{\ttm{A}} 
\underbrace{\frac{1}{2} \| \mathbf{y} - \mathcal{M}\!\left(\ttm{\hat{\rho}}\right) \|_2^2}_{=: f\left(\ttm{A}\right)}, 
 \quad
\text{s.t.} ~ \|\ttm{A}\|_{\mathrm{F}}^2 \leq 1, \quad \text{with } \ttm{\hat{\rho}} = \ttm{A} \ttm{A}^{\mathrm{H}}. 
\end{equation}
The map $\ten{M}$ provides compressed (trace-based) measurements of the underlying density matrix, $(\ten{M}(\ttm{\hat{\rho}}))_m = \operatorname{Tr}(\ttm{A}^{\mathrm{H}} \mat{E}_m  \ttm{A}^{\mathrm{H}} )$. Each measurement yields the expectation value of an observable and generally depends on all density matrix entries. 

\subsubsection*{DMRG-based approach to scalable QST}\label{subsubsec:dmrgqst}

We focus on measurement operators that admit low-rank TTM representations, as is the case for many important classes of measurement operators, including Pauli observables and a broad range of structured operators that can be well  approximated by sums of a relatively small number of Kronecker-product terms. Leveraging this structure, we first develop an efficient tensor-network contraction scheme for expectation-value evaluation, followed by a single-site DMRG optimization scheme for QST and a brief discussion of its two-site extension.

Consider measurement operators of the form, 
\begin{equation*}
\ttm{E}= \ubar{\mat{E}}^{(1)} \sktensor \ubar{\mat{E}}^{(2)} \sktensor \cdots \sktensor \ubar{\mat{E}}^{(N)}, 
\end{equation*}
where the cores satisfy $\ubar{\mat{E}}^{(n)} \in  \mathbb{C}^{R_{n\!-\!1}^{\scriptscriptstyle{E}} \cdot d \times  d \cdot R_{n}^{\scriptscriptstyle{E}}  }$.  There are various choices of observables; for example, for $N$-qubit systems ($d\!=\!2$), a common choice is the set of Pauli operators, which admit a TTM representation with TT-rank elements equal to one, i.e., $R_n^{\scriptscriptstyle{E}} \!=\! 1, \; \forall n$. In this case, the strong Kronecker product reduces to the standard one.

\paragraph{Efficient expectation values via tensor network contraction} \label{par:expectation}
The expectation value $\operatorname{Tr}(\ttm{A}^{\mathrm{H}} \ttm{E} \ttm{A})$ can be computed efficiently via corewise  contractions:
\begin{equation*}
\operatorname{Tr}\Bigg(\left(\ubar{\mat{A}}^{(1) \mathrm{H}} \sbullet \ubar{\mat{E}}^{(1)} \sbullet \ubar{\mat{A}}^{(1)} \right) \sktensor \cdots  \sktensor \left(\ubar{\mat{A}}^{(N) \mathrm{H}} \sbullet \ubar{\mat{E}}^{(N)} \sbullet \ubar{\mat{A}}^{(N)} \right)\Bigg).
\end{equation*}

\begin{figure}[htb]
	\centering
    \includegraphics[width=0.75\linewidth]{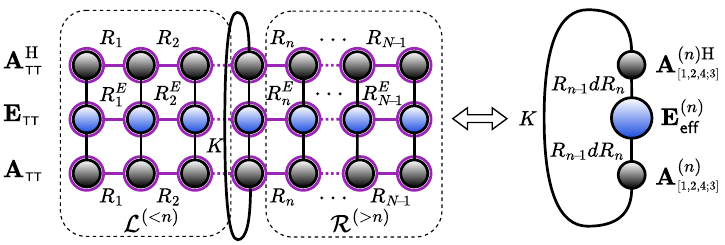}
	\caption{(Left) Core-wise contraction of the tensor network representing the expectation value computation, i.e., $\operatorname{Tr}(\ttm{A}^{\mathrm{H}} \,\ttm{E}\,  \ttm{A}).$ (Right) Equivalent tensor network representation of the same operation in terms of an effective operator, i.e., $\operatorname{Tr}(\mat{A}^{(n)\mathrm{H}}_{[1,2,4;3]}\, \eff{E}{n}\, \mat{A}^{(n)}_{[1,2,4;3]}),$ where $\eff{E}{n} = \mat{A}_{\neq n}^{\mathrm{H}} \,\ttm{E}\, \mat{A}_{\neq n} \in \mathbb{C}^{R_{n\!-\!1} d R_n \times R_{n\!-\!1} d R_n}$.}
\label{figTNexpectation} \vspace{-2mm}
\end{figure}

\Cref{figTNexpectation} visualizes the tensor network diagram for this operation: the overall contraction between the Block-TT parameterized density operator and the (TTM) measurement operators (left), alongside its equivalent effective-operator form (right).  In practice, the effective operator \( \eff{E}{n} \!=\! \mathbf{A}_{\neq n}^{\mathrm{H}} \, \ttm{E} \, \mathbf{A}_{\neq n} \) is never explicitly formed; instead, it is constructed via the so-called {\em left and right environments}, $\ten{L}^{(<n)}$ and $\ten{R}^{(>n)}$, which are computed recursively as shown in \eqref{eqn:env-recursion}. Once \(\ten{L}^{(<n)} \in \C^{R_{n\!-\!1} \times R^{\scriptscriptstyle E}_{n\!-\!1} \times R_{n\!-\!1}}\) and \(\ten{R}^{(>n)} \in \C^{R_n \times R^{\scriptscriptstyle E}_n \times R_n}\) are obtained, contracting the tensorized form $\ten{E}^{(n)} \in \C^{R^{\scriptscriptstyle E}_{n\!-\!1}\times d\times d\times R^{\scriptscriptstyle E}_n}$ of the local operator $\ubar{\mathbf{E}}^{(n)}$ with these environments yields:
\begin{equation}\label{eqn:effop1}
	\ten{Z} = \left(\ten{E}^{(n)} \bullet_{\scriptscriptstyle 1}^{\scriptscriptstyle 2}\ten{L}^{(<n)}\right)\bullet_{\scriptscriptstyle 3}^{\scriptscriptstyle 2}\ten{R}^{(>n)}; \quad  \eff{E}{n} = \mat{Z}_{[3,1,5\,;\,4,2,6]} \in \C^{R_{n\!-\!1} d R_{n} \times R_{n\!-\!1} d R_{n}}.
\end{equation}

\begin{figure}[htb]
			\centering
	       \includegraphics[width=0.9\linewidth]{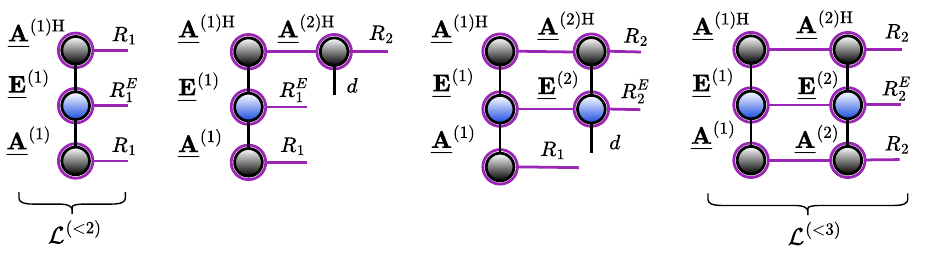}
			\caption{Optimal ordering of the tensor network contractions needed to compute the left and right environments recursively; see \eqref{eqn:env-recursion}.}
			\label{figleftenv}
		\end{figure}

\paragraph{Optimal contraction strategy}\label{app:optcontraction}
The tensor network in~\cref{figTNexpectation} is contracted efficiently by recursively constructing the left and right environments, with boundary conditions $\ten{L}^{(<1)} \!= \!1$ and $\ten{R}^{(>N)} \!= \!1$. As illustrated in \Cref{figleftenv}, $\ten{L}^{(<n)}$ is computed by successively contracting $\ten{L}^{(<n\!-\!1)}$ with $\ubar{\mathbf{A}}^{(n\!-\!1)\mathrm{H}}$, $\ubar{\mathbf{E}}^{(n\!-\!1)}$, and $\ubar{\mathbf{A}}^{(n\!-\!1)}$;  the right environments are obtained analogously. Specifically,  let 
	\begin{equation*}
		\mat{Z}^{(i)} = \ubar{\mat{A}}^{(i) \mathrm{H}} \sbullet \ubar{\mat{E}}^{(i)} \sbullet \ubar{\mat{A}}^{(i)}  \in \C^{R_{i\!-\!1} R^{\scriptscriptstyle E}_{i\!-\!1} R_{i\!-\!1}  \times R_{i} R^{\scriptscriptstyle E}_{i} R_{i}}, 
		\end{equation*}
	the left and right environment tensors are computed recursively by
	\begin{equation}
		\label{eqn:env-recursion}
		\begin{aligned}
			&\operatorname{vec}\!\left(\ten{L}^{(<i)}\right)^{\mathrm H}
			=
			\operatorname{vec}\!\left(\ten{L}^{(<i\!-\!1)}\right)^{\mathrm H}
			\mat{Z}^{(i\!-\!1)}
			\in
			\C^{1\times R_{i\!-\!1}R^{\scriptscriptstyle E}_{i\!-\!1}R_{i\!-\!1}}, \qquad i \!=\! 2, \cdots, n,  \\[3mm]
			& \operatorname{vec}\!\left(\ten{R}^{(>i)}\right)
			=
			\mat{Z}^{(i\!+\!1)} \operatorname{vec}\!\left(\ten{R}^{(>i\!+\!1)}\right)
			\in
			\C^{R_iR^{\scriptscriptstyle E}_iR_i \times 1},  \qquad i \!=\! n, \cdots, N\!-\!1. 
		\end{aligned}
	\end{equation}
Performing the contractions in the order shown in \Cref{figleftenv}, without explicitly forming $\mat{Z}^{(i)}$, each update of $\ten{L}^{(<i)}$ (or $\ten{R}^{(>i)}$) requires \(\mathcal{O}\!\left(d R^3 R^{\scriptscriptstyle E} \!+\! d^2 R^2 (R^{\scriptscriptstyle E})^2\right) \).  Thus, evaluating the expectation value in \Cref{figTNexpectation} costs
\(\mathcal{O}\!\left(
		\left(d R^3 R^{\scriptscriptstyle E} \!+\! d^2 R^2 (R^{\scriptscriptstyle E})^2\right)(N \!+\! K)
		\right) \),
		assuming \(R_n \!= \! R\) and \(R_n^{\scriptscriptstyle E} \!=\! R^{\scriptscriptstyle E}\). 
        
\paragraph{Single-site DMRG optimization}\label{par:dmrg-1}
Assuming $n$-orthogonality, we can convert the global optimization problem in \eqref{eqn:bttqst} into a set of linked small-scale optimization problems of the same form as 
\begin{equation}
		\label{eqn:bttqstlocal1}
		\scalebox{0.98}{$
		\begin{array}{ll}
			\displaystyle \min_{\ubar{\mat{A}}^{(n)}} \!&\! \underbrace{\frac{1}{2} \sum_{m=1}^M \left( y_m  - \operatorname{Tr}\left(\mat{A}^{(n)\mathrm{H}}_{[1,2,4;3]} \; \effm{E}{n}\; \mat{A}^{(n)}_{[1,2,4;3]}\right) \right)^2}_{ =: f\left(\ubar{\mat{A}}^{(n)} \right)},  \quad \text { s.t. }  \|\ubar{\mat{A}}^{(n)}\|_{\mathrm{F}}^2\leq 1, \\[5pt]  \text { with } & \effm{E}{n} \!=\! \mat{A}_{\neq n}^{\mathrm{H}} \; \ttmm{E} \; \mat{A}_{\neq n}, \quad \text { for }  n\!=\!1, 2, \cdots, N.
		\end{array}
		$}
\end{equation} 
Here, $\mat{A}^{(n)}_{[1,2,4;3]}\in \mathbb{C}^{R_{n\!-\!1}d R_n\times K}$ denotes the matrix unfolding of the $n$th core  $\ubar{\mat{A}}^{(n)}$. The unit-trace relaxation constraint reduces to a local constraint since 
\(
\|\ttm{A}\|_{\mathrm{F}}^2 = \|\ubar{\mat{A}}^{(n)}\|_{\mathrm{F}}^{2}
\),
thereby enabling efficient projection steps. The objective function above can be minimized, for example, using projected gradient descent (GD):
\(
\ubar{\mat{A}}^{(n)}_{i+1} = {\Pi}_{\ten{C}}\!\left(\ubar{\mat{A}}^{(n)}_{i} - \eta \nabla f\!\left(\ubar{\mat{A}}^{(n)}_{i}\right)\right),
\)
where $\Pi_{\mathcal{C}}$ denotes the projection onto the unit Frobenius ball $\mathcal{C} = \left\{\ubar{\mat{X}} \;\middle|\; \|\ubar{\mat{X}}\|_{\mathrm{F}}^2 \le 1 \right\}$, and $$\nabla f(\ubar{\mat{A}}^{(n)})=-2\sum_{m=1}^M \left( y_m  - \operatorname{Tr}\left(\mat{A}^{(n)\mathrm{H}}_{[1,2,4;3]} \; \effm{E}{n}\; \mat{A}^{(n)}_{[1,2,4;3]}\right) \right)\effm{E}{n}\; \mat{A}^{(n)\mathrm{H}}_{[1,2,4;3]} $$ denotes the Euclidean gradient  of the objective function $f$. We refer the reader to \cite{burer2003lrsdp, bhojanapalli2016lrsdp, park2018lowrank, kyrillidis2018provable, hu2019low} for further details on solving such problems. More advanced methods, such as (projected) Gauss--Newton, can also be employed for constrained nonlinear least-squares problems \cite{nocedal2006numerical}.  After solving the local subproblem, the block index $K$ is shifted to an adjacent core along with the orthogonality center, which becomes the next optimization target site. In this manner, all local sites are optimized sequentially across alternating sweeps until convergence. The overall procedure of the proposed method is outlined in \Cref{alg:ttqst-cs}.

\begin{algorithm}[htbp]
	\caption{Block-TT DMRG-I CS-QST}
	\label{alg:ttqst-cs}
	
	\begin{algorithmic}
		\REQUIRE Measurement data $\{y_m, \ttmm{E}\}_{m\!=\!1}^{M}$, initial  TT-rank $\left(R_0, \cdots, R_N\right)$, and $K$.
		\ENSURE $\ttm{A}$  such that $\ttm{\hat{\rho}} = \ttm{A} \ttm{A}^{\mathrm{H}}$. 
		
		\STATE \textbf{\scriptsize 1}: Initialize $\ttm{A}$ in Block$\!-\!1\!-\!$TT format   \ttc{$\ttm{A}$ is in right-orthogonal form}
		
		\STATE \textbf{\scriptsize 2}:  Set the left and right boundary environments as
		$\{\ten{L}_m^{(<1)} = \ten{R}_m^{(>N)} = 1\}_{m=1}^{M}$.
		
		\STATE \textbf{\scriptsize 3}: Compute  $\{ \ten{R}_m^{(>N\!-\!1)} \}_{m\!=\!1}^{M}, \cdots, \{ \ten{R}_m^{(>1)}\}_{m\!=\!1}^{M}$ recursively, as shown in \eqref{eqn:env-recursion}.

		\REPEAT 
		\STATE \ttc{left-to-right half sweep}
		
		\FOR{$n=1, 2, \cdots, N\!-\!1$}    
		
		\STATE \textbf{\scriptsize 4}: Compute $\{ \effm{E}{n}\}_{m\!=\!1}^{M}$ via $\{\ubar{\mat{E}}_{\, m}^{(n)}\}_{m\!=\!1}^{M}$, $\{ \ten{L}_m^{(<n)}\}_{m\!=\!1}^{M}$ and $\{ \ten{R}_m^{(>n)} \}_{m\!=\!1}^{M}$; see  \eqref{eqn:effop1}.
		
		\STATE \textbf{\scriptsize 5}: Minimize the local objective in~\eqref{eqn:bttqstlocal1} with respect to core $\ubar{\mat{A}}^{(n)}$.

		\STATE \textbf{\scriptsize 6}:  Transform Block$\!-\!n\!-\!$TT into  Block$\!-\!(n\!+\!1)\!-\!$TT, so that the updated core is left-orthogonal and hence $\ttm{A}$ is in \( \!(n\!+\!1)\!-\!\)orthogonal form, while updating the TT-rank. \label{line:moveK}
			
		\STATE  \ttc{Current $\ubar{\mat{A}}^{(n\!+\!1)}$ may serve as initialization for solving the $(n\!+\!1)$th core.}
		\STATE \textbf{\scriptsize 7}: Apply projection: $\ubar{\mat{A}}^{(n\!+\!1)} = \frac{\ubar{\mat{A}}^{(n+1)}}
{\max \left(1,~\|\ubar{\mat{A}}^{(n+1)}\|_F\right)}. ~$  \ttc{ensures $\| \ttm{A}\|_F\leq 1$}
		
	     \STATE \textbf{\scriptsize 8}:  Compute  $\{ \ten{L}_m^{(<n\!+\!1)}\}_{m\!=\!1}^{M}$ via \eqref{eqn:env-recursion}.

		\ENDFOR
		
		\STATE \ttc{right-to-left half sweep}
		
		\FOR {$n=N, N\!-\!1, \cdots, 2$}  
		\STATE \textbf{\scriptsize 9}: Compute $\{ \effm{E}{n}\}_{m\!=\!1}^{M}$ via $\{\ubar{\mat{E}}_{\, m}^{(n)}\}_{m\!=\!1}^{M}$, $\{ \ten{L}_m^{(<n)}\}_{m\!=\!1}^{M}$ and $\{ \ten{R}_m^{(>n)} \}_{m\!=\!1}^{M}$; see  \eqref{eqn:effop1}.
		
		\STATE \textbf{\scriptsize 10}: Minimize the local objective in~\eqref{eqn:bttqstlocal1} with respect to core $\ubar{\mat{A}}^{(n)}$.

		\STATE \textbf{\scriptsize 11}:  Transform Block$\!-\!n\!-\!$TT into  Block$\!-\!(n\!-\!1)\!-\!$TT, so that the updated core is right-orthogonal and hence $\ttm{A}$ is in \( \!(n\!-\!1)\!-\!\)orthogonal form, while updating the TT-rank.

	    \STATE  \ttc{Current $\ubar{\mat{A}}^{(n\!-\!1)}$ may serve as initialization for solving the $(n\!-\!1)$th core.}
			
		\STATE \textbf{\scriptsize 12}: Apply projection: $\ubar{\mat{A}}^{(n\!-\!1)} = \frac{\ubar{\mat{A}}^{(n-1)}}
{\max \left(1,~\|\ubar{\mat{A}}^{(n-1)}\|_F\right)}. ~$  \ttc{ensures $\| \ttm{A}\|_F\leq 1$}
		
		\STATE \textbf{\scriptsize 13}:  Compute  $\{ \ten{R}_m^{(>n\!-\!1)}\}_{m\!=\!1}^{M}$ via \eqref{eqn:env-recursion}.
		\ENDFOR
		
		\UNTIL{a stopping criterion is met.}
		\STATE  \textbf{return} $\ttm{A}$
	\end{algorithmic}
\end{algorithm}

\paragraph{Rank adaptation and block-index sweeping in DMRG-I}\label{par:rankadaptivity} 
During the left-to-right and right-to-left sweeps, the block index $K$ is sequentially shifted to the active optimization block alongside the orthogonality center; this ensures that the active core always carries the block parameter $K$, which enables automatic (block)-rank adaptation. More specifically, assume a left-to-right sweep, where $\ttm{A}$ is in Block$\!-\!n\!-\!$TT form and satisfies $n\!-\!$orthogonality. After optimizing the block-core $\ubar{\mat{A}}^{(n)} \in \mathbb{C}^{R_{n\!-\!1} \cdot d \times K \cdot R_n}$, $\ttm{A}$ is transformed to Block$\!-\!(n\!+\!1)\!-\!$TT form while also shifting the orthogonality center to $\ubar{\mat{A}}^{(n\!+\!1)}$ using the following procedure (Line~6 of ~\Cref{alg:ttqst-cs}).
Let the $\delta$-truncated SVD\footnote{SVD with singular values below tolerance $\delta$ discarded.} of the unfolding of the $n$th core be
\(
\left[\mat{U},\mat{S},\mat{V}\right]:=\mathtt{SVD}_{\delta}\left(\mat{A}^{(n)}_{[1,2;3,4]}\right),
\)
where $\mat{U}\in\mathbb{C}^{R_{n\!-\!1}d\times R_n^{\mathrm{new}}}.$
The TT-rank is updated as
\(
R_n = R_n^{\mathrm{new}}=\rank{\mat{U}},
\)
and $\mat{U}$ is reshaped into the new $n$th core with dimensions
$R_{n\!-\!1}\cdot d\times R_n^{\mathrm{new}}$. The remaining factor $\mat{SV}$, which carries the index $K$, is absorbed into the next core and appropriately permuted to preserve the Block-TT core structure.  Similarly, during the right-to-left sweep, $K$ and the orthogonality center are moved to the preceding core.

Notice that for $K\!=\!1$, the TT-rank cannot increase, since the updated rank is bounded by
\(
R_n^{\mathrm{new}}\leq \min(R_{n\!-\!1}d,KR_n)=\min(R_{n\!-\!1}d,R_n).
\) We show how this issue can be overcome within the \hyperref[par:dmrg-2]{two-site DMRG optimization} scheme, following previous works in the literature \cite{white1993density, schollwock2005dmrg, white2005density, khoromskij2010dmrgqtt, kressner2014btt, dolgov2014computation, lee2015btt}.

\paragraph{Two-site DMRG optimization}\label{par:dmrg-2}
The two-site DMRG-based QST method is a natural extension of the single-site DMRG-based QST method. In this approach, instead of optimizing one core at a time, two adjacent cores are combined into a larger core (a two-site core). This larger core is optimized and subsequently split back into two separate cores using truncated SVD. This method has been shown to achieve faster convergence per iteration \cite{schollwock2005dmrg, kressner2014btt, lee2015btt} and allow automatic rank adaptation, even when $K\!=\!1$. However, optimizing over a larger core tensor increases memory and computational requirements accordingly. The objective function can be expressed as:
\begin{equation}
	\label{eqn:bttqstlocal2}
	\scalebox{0.98}{$
	\begin{array}{ll}
		\displaystyle \min_{\ubar{\mat{A}}^{(n,n\!+\!1)}} \!&\! \underbrace{\!\frac{1}{2} \sum_{m=1}^M \left( y_m  \!-\! \operatorname{Tr}\left(\mat{A}^{(n, n\!+\!1)\mathrm{H}}_{[1,2,4;3]} \; \effm{E}{n, n\!+\!1} \; \mat{A}^{(n, n\!+\!1)}_{[1,2,4;3]}\right) \right)^2}_{ =: f\left(\ubar{\mat{A}}^{(n,n\!+\!1)} \right)},  ~ \text { s.t.  }  \|\ubar{\mat{A}}^{(n,n\!+\!1)}\|_{\mathrm{F}}^2\!\leq \!1, \\[5pt]
		\text {with}\!&\! \effm{E}{n, n\!+\!1} \!=\! \mat{A}_{\neq n, n\!+\!1}^{\mathrm{H}} \; \ttmm{E} \; \mat{A}_{\neq n, n\!+\!1}, \quad \text{ for } n\!=\!1, 2, \cdots, N\!-\!1. 
	\end{array}
	$}
\end{equation} 
Here, \(  \mat{A}^{(n,n\!+\!1)}_{[1,2,4;3]}\in \mathbb{C}^{R_{n\!-\!1}d^2 R_{n\!+\!1}\times K} \) denotes the matrix unfolding of the two-site core \(\ubar{\mat{A}}^{(n,n\!+\!1)} =  \ubar{\mat{A}}^{(n)} \sktensor \ubar{\mat{A}}^{(n\!+\!1)}\). The two-site effective operator is computed by contracting the tensorized two-site local operator $\ten{E}^{(n,n\!+\!1)}\in \C^{R^{\scriptscriptstyle E}_{n\!-\!1}\times d^2\times d^2\times R^{\scriptscriptstyle E}_{n\!+\!1}}$, obtained from the strong Kronecker product of the corresponding cores $\ubar{\mathbf{E}}^{(n)}\sktensor\ubar{\mathbf{E}}^{(n\!+\!1)}$, with the left and right environments, as follows:
\begin{equation}
	\label{eqn:effop2}
	\thinmuskip=1mu \medmuskip=2mu \thickmuskip=3mu
	\scalebox{0.92}{$
		\ten{Z} = \left(\ten{E}^{(n,n\!+\!1)}\bullet_{\scriptscriptstyle 1}^{\scriptscriptstyle 2}\ten{L}^{(<n)}\right)\bullet_{\scriptscriptstyle 3}^{\scriptscriptstyle 2}\ten{R}^{(>n\!+\!1)}, ~ \eff{E}{n, n\!+\!1} = \mat{Z}_{[3,1,5\,;\,4,2,6]} \in \C^{R_{n-1}d^2R_{n+1}\times R_{n-1}d^2R_{n+1}}.
		$}
\end{equation}
\paragraph{Rank adaptation and block-index sweeping in DMRG-II}\label{par:rankadaptivity2}
Similarly, during the left-to-right and right-to-left sweeps, the index $K$ is shifted between successive two-site cores so that it always remains attached to the active optimization block. Assuming a left-to-right sweep, after optimizing the two-site core $\ubar{\mat{A}}^{(n,n\!+\!1)}$, the index $K$ is shifted to the next position by splitting the supercore back into two cores via a truncated SVD. This simultaneously converts a Block$\!-\!n\!-\!$TT into a Block$\!-\!(n\!+\!1)\!-\!$TT and moves the orthogonality center to the $(n\!+\!1)$th core. Let $\mathcal{Z}^{(n,n\!+\!1)} \in \mathbb{C}^{R_{n\!-\!1} \times  d \times d \times K \times R_{n\!+\!1}}$ denote the tensor corresponding to $\ubar{\mat{A}}^{(n,n\!+\!1)}$, and consider the $\delta$-truncated SVD of its matrix unfolding: \(
\left[\mat{U},\mat{S},\mat{V}\right]:=\mathtt{SVD}_{\delta}\left(\mat{Z}^{(n,n\!+\!1)}_{[1,2;3,4,5]}\right),
\)
where $\mat{U}\in\mathbb{C}^{R_{n\!-\!1}d\times R_n^{\mathrm{new}}}.$
The TT-rank is updated as
\(
R_n = R_n^{\mathrm{new}}=\rank{\mat{U}}.
\)
The factor $\mat{U}$ is then reshaped into the new $n$th core of dimensions $R_{n\!-\!1}\cdot d\times R_n^{\mathrm{new}},$ while $\mat{SV}$ is reshaped into the new $(n\!+\!1)$th core of dimensions
$R_n^{\mathrm{new}} \cdot d\times  K \cdot R_{n\!+\!1}.$ This transfers the index $K$ and shifts the orthogonality center to the $(n\!+\!1)$th core. 

Notice that when $K\!=\!1$ the two-site DMRG allows rank adaptation since  the updated rank is now bounded by
\(
R_n^{\mathrm{new}}\leq \min(R_{n\!-\!1}d, dKR_{n\!+\!1})=\min(R_{n\!-\!1}d,dR_{n\!+\!1}).
\) In contrast to single-site DMRG, the two-site DMRG scheme allows for a larger increase in the TT-rank when needed. However, this added flexibility comes at the cost of larger local optimization problems, which may slow down the overall optimization per iteration. The overall procedure is summarized  in \Cref{alg2:ttqst-cs}.

\begin{algorithm}[htbp]
	\caption{Block-TT DMRG-II CS-QST}
	\label{alg2:ttqst-cs}
	
\begin{algorithmic}
	\REQUIRE Measurement data $\{y_m, \ttmm{E}\}_{m\!=\!1}^{M}$, initial  TT-rank $\left(R_0, \cdots, R_N\right)$, and $K$.
	\ENSURE $\ttm{A}$  such that $\ttm{\hat{\rho}} = \ttm{A}\ttm{A}^{\mathrm{H}}$. 
	
	\STATE \textbf{\scriptsize 1}: Initialize $\ttm{A}$ in Block$\!-\!1\!-\!$TT format   \ttc{$\ttm{A}$ is in right-orthogonal form}
	
	\STATE \textbf{\scriptsize 2}:  Set the left and right boundary environments as
	$\{\ten{L}_m^{(<1)} = \ten{R}_m^{(>N)} = 1\}_{m=1}^{M}$.
	
	\STATE \textbf{\scriptsize 3}: Compute  $\{ \ten{R}_m^{(>N\!-\!1)} \}_{m\!=\!1}^{M}, \cdots, \{ \ten{R}_m^{(>1)}\}_{m\!=\!1}^{M}$ recursively,  as shown in \eqref{eqn:env-recursion}.

	\REPEAT 
	\STATE \ttc{left-to-right half sweep}
	
	\FOR{$n=1, 2, \cdots, N\!-\!1$}     
	
	\STATE \textbf{\scriptsize 4}: Compute \(\{ \effm{E}{n, n\!+\!1}\}_{m\!=\!1}^{M}\) via  \(\{\ubar{\mathbf{E}}_{\, m}^{(n)} \sktensor \ubar{\mathbf{E}}_{\, m}^{(n\!+\!1)}\}_{m\!=\!1}^{M}\), \(\{ \ten{L}_m^{(<n)}\}_{m\!=\!1}^{M}\) and \(\{ \ten{R}_m^{(>n\!+\!1)} \}_{m\!=\!1}^{M}\), as shown in \eqref{eqn:effop2}.

	\STATE \textbf{\scriptsize 5}: Minimize the local objective in~\eqref{eqn:bttqstlocal2} with respect to the two-site core $\ubar{\mat{A}}^{(n)}$.

	\STATE \textbf{\scriptsize 6}:  Split the updated two-site core $\ubar{\mat{A}}^{(n,n\!+\!1)}$ via a truncated SVD, updating the TT-rank, shifting the orthogonality center, and converting the Block$\!-\!n\!-\!$TT into a Block$\!-\!(n\!+\!1)\!-\!$TT.

	\STATE \textbf{\scriptsize 7}: Apply projection:  $\ubar{\mat{A}}^{(n\!+\!1)} = \frac{\ubar{\mat{A}}^{(n+1)}}
{\max \left(1,~\|\ubar{\mat{A}}^{(n+1)}\|_F\right)}. ~$  \ttc{ensures $\| \ttm{A}\|_F\leq 1$}
	
	\STATE \textbf{\scriptsize 8}:  Compute  $\{ \ten{L}_m^{(<n\!+\!1)}\}_{m\!=\!1}^{M}$ via \eqref{eqn:env-recursion}.
	
	\ENDFOR
	
	\STATE \ttc{right-to-left half sweep}
	
	\FOR {$n=N, N\!-\!1, \cdots, 2$}  
	
	\STATE Carry out the  right-to-left half-sweep similarly. 
	
	\ENDFOR
	
	\UNTIL{a stopping criterion is met.}
	\STATE  \textbf{return} $\ttm{A}$
\end{algorithmic}
\end{algorithm}

\paragraph{Overlapping  measurement operators for efficient environment caching}\label{par:overlap}

Consider a family of measurement operators of the form, 
\begin{equation*}
\ttm{E} = \underbrace{ \ubar{\mat{E}}^{(1)} \sktensor \cdots \sktensor \ubar{\mat{E}}^{(n_1-1)}}_{\text{fixed left}} \sktensor \underbrace{ \ubar{\mat{E}}^{(n_1)} \sktensor \cdots \sktensor \ubar{\mat{E}}^{(n_2)}}_{\text{variable block}} \sktensor \underbrace{ \ubar{\mat{E}}^{(n_2+1)} \sktensor \cdots \sktensor \ubar{\mat{E}}^{(N)}}_{\text{fixed right}}
\end{equation*}

\noindent where the left cores $\{\ubar{\mat{E}}^{(i)}\}_{i<n_1}$ and the right cores $\{\ubar{\mat{E}}^{(i)}\}_{i>n_2}$ are the same for the whole family, while the operator cores inside the block $[n_1, n_1\!+\!1, \cdots,  n_2]$ vary (the single‑qudit case corresponds to $n_1\!=\!n_2$).   Here, we pre‑contract the fixed left and right parts into environment tensors $\ten{L}^{(<n_1)}$ and $\ten{R}^{(>n_2)}$. The expectation value of any member in this family then reduces to a contraction over only the variable block tensors. This enables efficient environment caching and expectation value evaluation.

To perform QST with this scheme, one can optimize the cores of $\ttm{A}$ over the measurements, with the variable block shifted left or right in a sliding-window fashion. Only the cores corresponding to the variable block can be optimized. This allows local POVMs to be measured on a subset of qubits while the remaining qubits are kept in a fixed basis. The approach is particularly useful for one-dimensional systems in which correlations are primarily restricted to neighbouring qubits. Similar ideas have been proposed, including efficient MPS tomography protocols for many-body systems based on local reductions (i.e., partial traces of the density matrix) \cite{o2016efficient, lanyon2017efficient, guo2024quantum,  tang2025sketch}. \par

\begin{figure}[htb]
	\centering
	\includegraphics[width=0.65\linewidth]{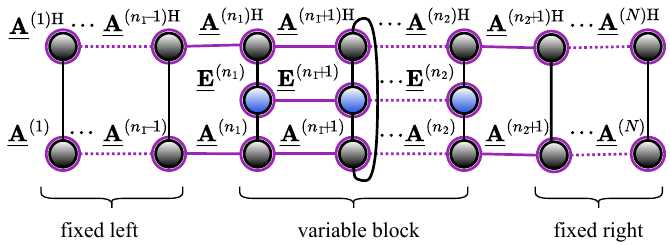}
	\caption{Expectation value evaluation on a local subsystem, equal to that obtained from the reduced density matrix after tracing out the fixed   parts.}
	\label{figTNexpectationReduced}
	\vspace{-3mm}
\end{figure}

Furthermore, if the left and right cores (the fixed parts) of the measurement operators are identity matrices, i.e.,  $\{\ubar{\mat{E}}^{(i)} = [\mat{I}_{d}] \in \mathbb{C}^{d \times d}\}_{i<n_1 \text{ or }  i>n_2},$  evaluating the full-state trace reduces to evaluating the trace on the reduced density matrix after tracing out fixed subsystems, as shown in \Cref{figTNexpectationReduced}.

\unskip
Note that our approach differs fundamentally from the MPS tomography methods proposed in \cite{o2016efficient, lanyon2017efficient, guo2024quantum}, which aim to reconstruct the full density matrix from exclusively local reduced density matrices. In contrast, our method estimates expectation values from (full) trace-based measurements and leverages a Block-TT parameterization that is directly applicable to mixed states.

\paragraph{Implementation details}
\Cref{alg:ttqst-cs,alg2:ttqst-cs} are implemented in the PyTorch framework, using an L-BFGS algorithm to solve the local optimization problems \eqref{eqn:bttqstlocal1} and \eqref{eqn:bttqstlocal2}. A link to the repository will be made available upon publication. 

\subsection{Computational complexity}
To assess the computational complexity of \Cref{alg:ttqst-cs,alg2:ttqst-cs}, we consider an $N$-qudit system and assume uniform ranks for the Block-TT tensor $\ttm{A}$ and the measurement operators $\ttm{E}$, i.e., $R_n=R$ and $R_n^{\scriptscriptstyle E}=R^{\scriptscriptstyle E}$. We report the complexity per sweep, assuming that the ranks and block index $K$ remain bounded during optimization. The dominant computational cost arises from (i) constructing the effective operators for the $M$ measurements (Line~4) and 
(ii)~solving the reduced optimization problem (Line~5), assuming $N_{\mathrm{eval}}$ evaluations of the objective function and its gradient are required to satisfy the stopping criterion. The corresponding complexities are summarized in \Cref{tab:dmrg_steps}. The costs associated with the orthogonalization and truncation steps (Lines~6 and~7) are negligible, as these operations act only on one-site or two-site cores and are independent of $M$.
Moreover, owing to the efficient contraction strategy described in \Cref{app:optcontraction}, the update of the left and right environments (Line~8) does not constitute the dominant computational cost.

\begin{table}[htbp]
\centering
\caption{Computational complexity per sweep (flop) of \Cref{alg:ttqst-cs,alg2:ttqst-cs}, where \(N_{\mathrm{eval}}\) denotes the number of objective and gradient evaluations required by the chosen local solver.}
\label{tab:dmrg_steps}
\resizebox{\linewidth}{!}{%
\begin{tabular}{@{} l c c @{}}
\toprule
\textbf{Dominant Terms} & \textbf{DMRG-I} & \textbf{DMRG-II} \\
\midrule
\begin{tabular}[c]{@{}l@{}}Effective operators (line 4)\end{tabular} & 
$\mathcal{O}\!\left(NM\left(d^2R^4R^{\scriptscriptstyle E} + d^2R^2(R^{\scriptscriptstyle E})^2\right)\right)$ & 
$\mathcal{O}\!\left(NM\left(d^4R^4R^{\scriptscriptstyle E} + d^4R^2(R^{\scriptscriptstyle E})^2\right)\right)$ \\
\addlinespace[4pt]
Solving local problem (line 5) & $\mathcal{O}\!\left(N N_{\mathrm{eval}}Md^2R^4K\right)$ & $\mathcal{O}\!\left(N N_{\mathrm{eval}}Md^4R^4K\right)$ \\
\bottomrule
\end{tabular}
}
\end{table}

Per measurement, the construction of the effective operator at a given site (DMRG-I) or site pair (DMRG-II) requires $\mathcal{O}\left(d^2R^4R^{\scriptscriptstyle E}+d^2R^2(R^{\scriptscriptstyle E})^2\right)$ operations for DMRG-I (\Cref{eqn:effop1}) and $\mathcal{O}\left(d^4R^4R^{\scriptscriptstyle E}+d^4R^2(R^{\scriptscriptstyle E})^2\right)$ for DMRG-II  (\Cref{eqn:effop2}). Furthermore, each evaluation of the local objective function and its gradient incurs a cost of $\mathcal{O}\left(Md^2R^4K\right)$ for DMRG-I and $\mathcal{O}\left(Md^4R^4K\right)$ for DMRG-II.

\section{Numerical experiments} \label{sec:exps}

This section evaluates the performance of the proposed methods across the following three experimental settings:
\begin{itemize}
	\item First, we show the accuracy and computational efficiency of the proposed approaches by comparing them with standard low-rank baselines: the rank-minimization formulation \eqref{eqn:sdpQST}, solved using the splitting conic solver in CVXPY \cite{cvxpy}, and the error-minimization approach \eqref{eqn:BMQST} based on the Burer--Monteiro factorization \cite{burer2003lrsdp}, as implemented in \cite{kyrillidis2018provable}. Performance is assessed using fidelity, \( \left(\operatorname{Tr}\!\left(\sqrt{\sqrt{\rho}\hat{\rho}\sqrt{\rho}}\right)\right)^2, \) and trace distance, \( \frac{1}{2}\|\rho-\hat{\rho}\|_1, \) which quantify (overlap) similarity and distinguishability, respectively.
	
	\item Second, we compare the two DMRG methods in terms of rank adaptivity, computational time, and reconstruction accuracy as the number of qubits increases. Here, performance is primarily evaluated using the Frobenius norm between the ground-truth and reconstructed states, as well as the accuracy of expectation values computed for unseen (test) measurement operators.
	
	\item Finally, we showcase that overlapping operators enable large-scale QST for structured states whose correlations are predominantly local, i.e., restricted to neighbouring qubits. As an illustrative example, we consider a structured MPS state and evaluate reconstruction accuracy in terms of fidelity.
\end{itemize}
The experiments were performed on an HP EliteBook 845 G8 Notebook PC equipped with an AMD Ryzen 7 PRO 5850U processor and 32 GB of RAM.

\subsection{Accuracy and computational efficiency}\label{acc} 
A random complex mixed state of an $N$-qubit system is generated by sampling a Block-TT tensor \(\ttm{A}\) with $K=2$ and a TT-rank with maximum entry $3$. The entries of each core tensor are complex numbers whose real and imaginary components are drawn independently from a standard normal distribution. After normalizing \(\ttm{A}\)  to unit norm, the corresponding TTM density operator is constructed as  \(\ttm{\rho}=\ttm{A}\ttm{A}^{\mathrm H}\).

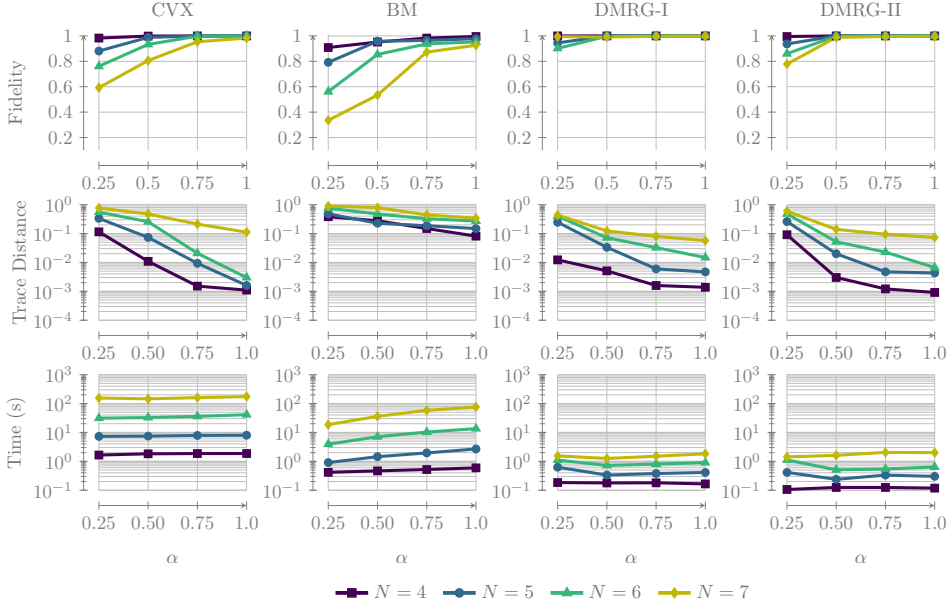
\begin{figure}[htbp]
    \centering
    \input{figures/mat_comparison.tex}
     \caption{Baseline comparison of reconstruction fidelity, trace distance, and runtime versus sampling rate $\alpha$ for system sizes $N \in \{4,5,6,7\}$. The proposed DMRG algorithms consistently outperform standard baselines, achieving near-perfect fidelity ($\ge 0.99$ for $\alpha \ge 0.5$) and over $10\times$ speedup at $N=7$.}
    \label{fig:baseline}
\end{figure}

The number of qubits $N$ is varied from 4 to 7. For each system size, local SIC-POVM measurements are performed, where each $N$-qubit measurement operator is formed by taking the Kronecker product of $N$ single-qubit SIC-POVM elements. The total number of measurement operators scales as $M = \alpha \cdot P \log(N)$, where $P= \sum_{n=1}^N  R^2_{n\!-\!1} d^2  R^2_n$ denotes the parameters of the equivalent MPO representation \cite{qin2024quantum, qin2026QSTreview}. This logarithmic scaling is directly motivated by the logarithm of the covering number ($\epsilon$-entropy), which quantifies the metric complexity of the underlying state class \cite{qin2024quantum, qin2026QSTreview} and provides a useful basis for designing QST evaluation experiments. 
The sampling factor $\alpha$ is varied from 0.25 to 1.0, and Gaussian noise is added to the measurements with signal-to-noise ratio ($\mathrm{SNR}$) equal to \qty{60}{\decibel}. The proposed algorithms (\Cref{alg:ttqst-cs,alg2:ttqst-cs}) are executed alongside reference methods, including a convex optimization approach (CVX) \cite{gross2010csqst, cvxpy} and a low-rank Burer--Monteiro solver (BM) \cite{kyrillidis2018provable}. For the DMRG algorithms, the relative objective function tolerance is set to
$10^{-4}$, chosen below the expected $10^{-3}$ relative noise level corresponding
to an SNR of $60\,\mathrm{dB}$, with a maximum of 5 sweeps. Reference methods use the same tolerance, terminating after a maximum of 400 iterations for CVX and 2000 for BM if convergence is not reached. The median fidelity and trace distance across 10 trials are reported in \Cref{fig:baseline}. 

The proposed DMRG algorithms consistently outperform CVX and BM in both accuracy and efficiency. By exploiting the TTM structure, DMRG-I and DMRG-II achieve near-perfect fidelity ($\ge 0.99$ for $\alpha \ge 0.5$) and significantly lower trace distance error, while running over an order of magnitude faster. Across all methods, increasing the sampling factor $\alpha$ enhances reconstruction accuracy with minor computational cost. Notably, at low sampling rates ($\alpha = 0.25$), DMRG-I slightly outperforms DMRG-II because limited data allows smaller single-core subproblems to be solved more accurately than larger two-core ones; as $\alpha$ increases, DMRG-II achieves superior performance. Finally, performance degradation at larger $N$ is primarily due to fixed sweep and iteration budgets, which can be resolved by scaling maximum iterations with system size.

\subsection{Rank adaptivity and scalability}\label{rankadap} 
In the first experiment, the rank adaptivity and convergence of the proposed DMRG-I and DMRG-II algorithms are evaluated. A random $N \!=\! 7$-qubit ground truth state is generated in Block-TT form with TT-rank $(1, 2, 3, 9, 10, 4, 2, 1)$ and $K \!=\! 2$, and local SIC-POVM measurements are performed with $M \!=\! 0.5 \cdot P \log(N)$ measurements. Gaussian noise is added to the measurements with an $\mathrm{SNR} = 60\text{~dB}$. QST is then performed using both methods, recording the objective loss per site in DMRG-I and per pair of consecutive sites in DMRG-II. The maximum number of sweeps is set to $3$ in both methods, and the relative objective function tolerance is set to $10^{-4}$ (according to the noise level). The same random initialization is used in both methods, with the TT-rank initialized to \( (1, 1, \ldots, 1) \).  During optimization, the maximum element of the TT-rank, $R_{\mathrm{max}}$, is recorded.
Median results are computed over $10$ independent trials and are plotted in \Cref{fig:rankadap}. It is observed that DMRG-II achieves faster convergence within one full sweep, whereas DMRG-I achieves similar convergence in two sweeps. As shown in the inset plot of \Cref{fig:rankadap}, ranks adapt faster in DMRG-II, whereas rank growth in DMRG-I is $\mathcal{O}(K)$. Thus, for small $K$, the rank adapts slowly, and more sweeps are required to converge.
\begin{figure}[htbp]
    \centering
    \input{figures/loss_sweep}
    \caption{Convergence and rank adaptivity comparison between DMRG-I and DMRG-II ($N \!=\! 7$, true $R_{\mathrm{max}} \!=\! 10$). The main plot shows the median loss across visited sites/site-pairs, with DMRG-II converging within a single sweep. The inset shows that DMRG-II reaches the target maximum rank ($R_{\mathrm{max}} \!=\! 10$, dotted line) by sweep 1, whereas DMRG-I adapts more gradually over two sweeps.}
\label{fig:rankadap}
    \label{fig:rankadap}
\end{figure}
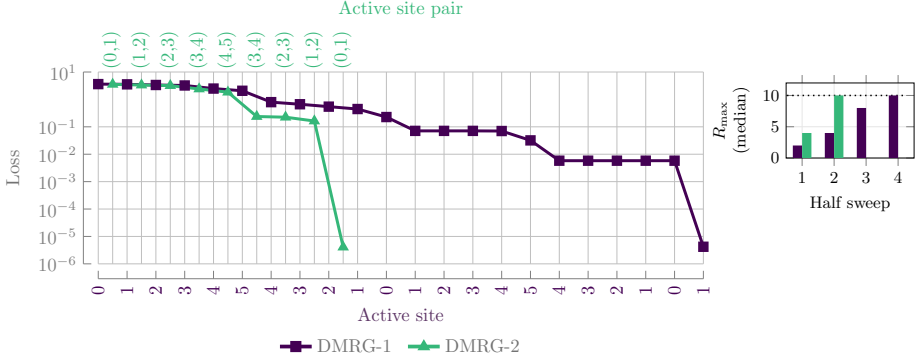

\begin{figure}[htbp]
    \centering
    \input{figures/scalability}
    \caption{Scalability comparison between DMRG-I and DMRG-II across qubit counts $N$: Frobenius error (left), test prediction error (middle), and computation time (right). While DMRG-II achieves higher accuracy and speed for small systems ($N \le 6$), its accuracy advantage vanishes for larger $N$, where its runtime increases significantly faster than DMRG-I.}
    \label{fig:scalability}
\end{figure}
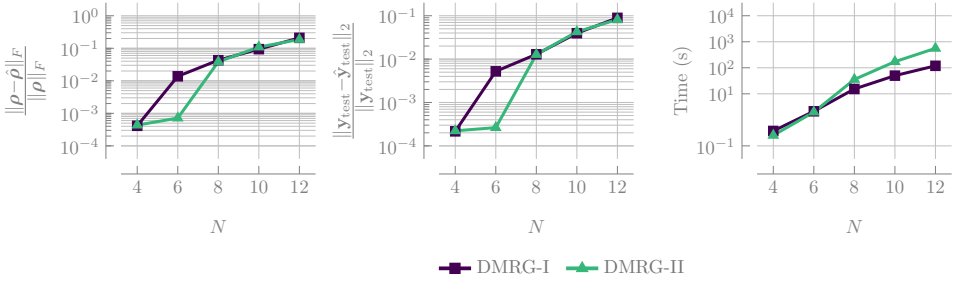

Next, the number of qubits $N$ is varied from $4$ to $12$. As in previous experiments, the ground-truth state is generated, local SIC-POVM measurements are performed, and outcomes are corrupted with an $\mathrm{SNR} = 60\text{ dB}$. QST is executed with a relative objective tolerance of $10^{-4}$ and up to $10$ sweeps. Three metrics are evaluated: relative Frobenius reconstruction error, relative norm distance between true and recovered expectation values on a test set ($\frac{1}{4}$ of the training set size), and computation time to satisfy the stopping criterion. Median results over $10$ independent trials are plotted in \Cref{fig:scalability}. While higher accuracy is achieved by DMRG-II for smaller systems, the accuracy gap between DMRG-I and DMRG-II vanishes as system size increases. A similar crossover is observed in computation time: DMRG-II converges slightly faster for smaller systems, but becomes slower as $N$ grows. This slowdown occurs because faster TT-rank growth in DMRG-II increases the cost of solving two-site cores and performing SVDs. Furthermore, a general increase in reconstruction error is observed for larger systems due to the fixed sweep count; maintaining convergence requires both iterations and measurement counts to scale strongly with $N$.

\begin{figure}[htbp]
	\centering
	\input{figures/K_approx}
	\caption{Reconstruction fidelity (left) and computation time (right) versus qubit count $N$ across model capacities $K$. Accuracy peaks when $K = K_{\mathrm{true}}$, and runtime increases with system size $N$.}
	\label{fig:K_approx}
\end{figure}
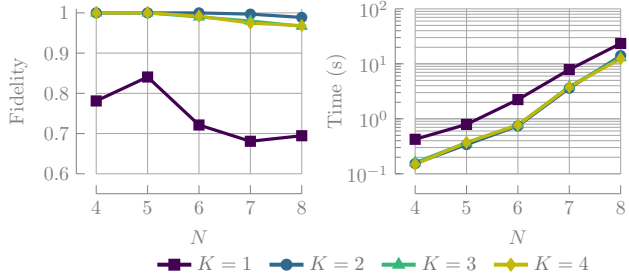

Finally, the impact of the block parameter $K$ on estimation accuracy and runtime is evaluated to demonstrate performance trade-offs when tuning model capacity. Random $N$-qubit ground-truth states ($N \in \{4, 5, 6, 7, 8\}$) are generated in Block-TT form with TT-ranks $(1, 3, \dots, 3, 1)$ and fixed true capacity $K_{\mathrm{true}} = 2$. Local SIC-POVM measurements are performed with $M = 0.2 \cdot P \log(N)$, and the measurement outcomes are corrupted with additive noise corresponding to an $\mathrm{SNR} = 60~\text{dB}$. DMRG-II is then used to reconstruct the state for $K \in \{1, 2, 3, 4\}$. The maximum sweep count is set to $4$ with a relative objective tolerance of $10^{-5}$. Median fidelity and computation time over $20$ independent trials are plotted in \Cref{fig:K_approx}. As $K$ increases from $1$ toward $K_{\mathrm{true}}$, state fidelity increases significantly, achieving near-unit fidelity across all system sizes. Smaller $K$ values require slightly longer optimization times; however, larger values converge faster. Notably, at $K = 2$, reconstruction achieves the best fidelity, whereas when $K$ exceeds the true value, fidelity decreases slightly because extra parameters tend to fit the noise.

\subsection{Exploiting overlapping measurements for large-scale QST}\label{overlap}
To show that overlapping localized measurements enable large-scale QST, we prepare a target state using the single-site DMRG algorithm implemented in PyTeNet \cite{mendl2018pytenet}. Specifically, we obtain the ground state of an $N=30$ site 1D transverse-field Ising model Hamiltonian:
\vspace{-2mm}
\begin{equation}
\mat{H} = -J \sum_{n=1}^{N-1} \bm{\sigma}_z^{(n)} \bm{\sigma}_z^{(n+1)}  - g \sum_{n=1}^N \bm{\sigma}_x^{(n)},
\end{equation}
with parameters $J = 1.0$ and $g = 2.0$, where $\bm{\sigma}_{\alpha}^{(n)} = \mat{I}_2^{\otimes n\!-\!1} \otimes \bm{\sigma}_{\alpha} \otimes \mat{I}_2^{\otimes N\!-\!n}$ is the $2^N \times 2^N$ operator obtained by placing the single-site Pauli matrix $\bm{\sigma}_{\alpha}$ at site $n$, with $\alpha \in \{z,x\}$. The target state is obtained as an MPS/TT with the maximum element of the TT-rank $R_{\mathrm{max}} = 6$.

We construct rank-1 TT measurement operators $\ttm{E}$ localized to a spatial window $[n_1, n_2]$: \vspace{-2mm}
\begin{equation*}
	\ttm{E} = \underbrace{ \mat{I}_2^{\otimes n_1\!-\!1}}_{\text{fixed}} \otimes \underbrace{\Big(\mat{E}(\mathbf{n}_{n_1}) \otimes \cdots \otimes \mat{E}(\mathbf{n}_{n_2}) \Big)}_{\text{variable block}} \otimes \underbrace{\mat{I}_2^{\otimes N\!-\!n_2}}_{\text{fixed}},
\end{equation*}
where $\mat{E}(\mathbf{n}_i) = \tfrac{1}{2}\left(\mat{I}_2 + n_{ix}\bm\sigma_x + n_{iy}\bm\sigma_y + n_{iz}\bm\sigma_z\right) $, with $\mathbf{n}_i = (n_{ix}, n_{iy}, n_{iz})^\top$ uniformly distributed over the surface of the Bloch sphere (solid angle $4\pi$) via  \(n_{ix} = \sin\alpha_i\cos\varphi_i \), \(\ n_{iy} = \sin\alpha_i\sin\varphi_i\),  \(\ n_{iz} = \cos\alpha_i,\)  with  \(\cos\alpha_i \sim \mathrm{Uniform}[-1,1] \), \( \varphi_i \sim \mathrm{Uniform}[0, 2\pi) \). Notice that the pair $\{\mat{E}(\mathbf{n}_i),  \mat{E}(-\mathbf{n}_i)\}$ forms a valid two-outcome projective POVM satisfying $\mat{E}(\mathbf{n}_i) + \mat{E}(-\mathbf{n}_i) = \mat{I}_2$.

Measurement operators are generated by sweeping a 1D window of width
$w = n_2 \!-\! n_1 \!+\! 1$ across the $N$-qubit chain with stride $s$.
At each window position $[n_1, n_2]$, a batch of product measurement
operators is sampled: the local Bloch vectors
$\{\vec{n}_i\}_{i=n_1}^{n_2}$ are drawn independently and
uniformly at random, while every site outside
the window is assigned the identity $\mat{I}_2$.
Shifting the window by $s$ sites gives a spatial overlap of
$o = w \!-\! s$ sites between consecutive windows ($s \leq w$), or
equivalently a relative overlap of $(1 - s/w)\times 100\%$. We study four window--stride configurations,
\( (w,\,s)\;\in\;\{(3,3),\,(6,3),\,(4,2),\,(4,1)\}, \)
with overlap percentages of $0\%$, $50\%$, $50\%$, and $75\%$,
respectively.
For accurate reconstruction, the window width and overlap must be
large enough to jointly capture all site pairs that are significantly
correlated in the target ground state.

In the first experiment, the total number of measurement operators is fixed at $2400$ for every
configuration.
The number of window positions is
\(N_p = \left\lfloor\frac{N - w}{s}\right\rfloor + 1, \)
so each position receives approximately
$\lfloor 2400 / N_p \rfloor$ operators.
After generation, the expectation values are corrupted by additive
white noise at $\mathrm{SNR} = 60\,\mathrm{dB}$.
The DMRG-I algorithm then reconstructs the quantum state.
Left and right environment tensors for the identity operators outside
the active window are computed once and cached; since consecutive
window positions share identical operators on all sites outside their
union, the corresponding environment tensors are reused without
recomputation. Table~\ref{tab:fixed_total} shows that, under the same total
measurement budget, the choice of $(w, s)$ leads to substantially
different reconstruction fidelity and computation time.

\begin{table}[h]
	\centering
	\caption{Reconstruction fidelity and execution time (mean $\pm$ std
        over 10 trials) for a fixed total budget of $2400$ measurement
        operators. Zero overlap yields substantially lower fidelity,
        while sufficient overlap enables accurate reconstruction}
	\label{tab:fixed_total}
	\begin{tabular}{cccccc}
		\toprule
		$s$ & $w$ & Overlap & $N_p$ & Fidelity & Time (s) \\
		\midrule
		3 & 3 & $0\%$  & 10 & $0.7026 \pm 0.0165$ & $146.04 \pm 31.23$ \\
		3 & 6 & $50\%$ &  9 & $0.9616 \pm 0.0171$ & $227.99 \pm 42.22$ \\
		2 & 4 & $50\%$ & 14 & $0.9750 \pm 0.0032$ & $208.04 \pm 14.59$ \\
		1 & 4 & $75\%$ & 27 & $0.9840 \pm 0.0037$ & $221.95 \pm  8.26$ \\
		\bottomrule
	\end{tabular}
\end{table}

In the second experiment, each window position is assigned exactly $100$ measurement operators,
so the total count $100 N_p$ varies across configurations.
Under the same four configurations, we compare
reconstruction fidelity and computation time.
Table~\ref{tab:fixed_per_pos} summarizes the results.

\begin{table}[h]
	\centering
	\caption{Reconstruction fidelity and execution time (mean $\pm$ std
        over 10 trials) for a fixed $100$ operators per window position.
        By tuning window width and stride, one can trade off between
        reconstruction fidelity and computational cost, provided the
        total measurement budget remains sufficient.}
	\label{tab:fixed_per_pos}
	\begin{tabular}{cccccc}
		\toprule
		$s$ & $w$ & Overlap & $N_p$ & Fidelity & Time (s) \\
		\midrule
		3 & 3 & $0\%$  & 10 & $0.6941 \pm 0.0174$ & $ 64.95 \pm  6.31$ \\
		3 & 6 & $50\%$ &  9 & $0.7990 \pm 0.0349$ & $125.77 \pm 12.92$ \\
		2 & 4 & $50\%$ & 14 & $0.9697 \pm 0.0038$ & $110.09 \pm  7.03$ \\
		1 & 4 & $75\%$ & 27 & $0.9854 \pm 0.0025$ & $196.21 \pm  3.91$ \\
		\bottomrule
	\end{tabular}
\end{table}

By tuning the window width and stride, one can trade off reconstruction
fidelity against computational cost; however, sufficient spatial overlap
and total measurement budget must both be maintained to ensure that all
physically relevant correlations of the target state are adequately
captured.
\section{Conclusion} \label{sec:con}

We introduced a Block-TT Burer--Monteiro framework for quantum state tomography of high-dimensional mixed quantum states. The proposed parameterization generalizes the Burer--Monteiro factorization to high-dimensional density matrices with a parameter count scaling linearly in the number of qudits, making it both memory- and computation-efficient. Building on this framework, we developed rank-adaptive single-site and two-site DMRG algorithms for learning low-rank mixed states. The resulting approach produces physically valid states by construction; thereby avoiding explicit positivity constraints. We further derived efficient tensor-network contraction schemes for local core optimization and introduced overlapping measurement operators together with effective environment caching strategies to enable scalable tomography of large structured quantum systems. Numerical experiments demonstrate that the proposed framework can accurately reconstruct low-rank quantum states from limited measurements while offering substantial reductions in storage and computational requirements.


\bibliographystyle{siamplain}
\bibliography{ref}
\end{document}

%% file: figures/mat_comparison.tex
\begin{tikzpicture}[scale=0.72]
	
	\begin{groupplot}[
		group style={group size=4 by 3, horizontal sep=1.5cm, vertical sep=1cm},
		width=0.33\textwidth,
		grid=both,
  	  legend style={
			at={(-1.65,-0.75)},
			anchor=north,
			legend columns=4,
			draw=none,
			fill=none,
			text=black!10!gray,
			font=\normalsize,
			/tikz/every even column/.append style={column sep=0.2cm}
		},
	    legend image post style={line width=2pt},
		every axis x label/.append style={
			at={(axis description cs:0.5,-0.5)},
			anchor=north, color=black!10!gray,
		},
		every axis y label/.append style={
			at={(axis description cs:-0.45,0.5)},
			anchor=south,
			color=black!10!gray,
		},
		x axis line style={yshift=-8pt,-,color=gray},
		y axis line style={xshift=-8pt,-,color=gray},
		xticklabel style={yshift=-8pt,color=gray, font=\normalsize},
		yticklabel style={xshift=-8pt,color=gray, font=\normalsize},
		xtick style={yshift=-8pt, line width=0.3pt, font=\normalsize},
		ytick style={xshift=-8pt, line width=0.3pt, font=\normalsize},
		xtick align=inside,
		ytick align=inside,
	]
		
		
		\nextgroupplot[
		ylabel={Fidelity},
		title={\color{black!10!gray}  CVX},
		axis lines=left,
		xtick={0.25, 0.5, 0.75, 1},
		xticklabels={0.25, 0.5, 0.75, 1},
		ymin=0.1, ymax=1.01,
		]
		\addplot[ultra thick, virpurple, mark=square*, mark size=1.5pt] coordinates {(0.25, 0.9829861) (0.50, 0.99875665) (0.75, 0.9999827) (1.0, 0.99992853)};
		\addplot[ultra thick, virblue, mark=*, mark size=1.5pt] coordinates {(0.25, 0.8809421) (0.50, 0.98844784) (0.75, 0.9991794) (1.0, 0.99993217)};
		\addplot[ultra thick, virgreen, mark=triangle*, mark size=1.5pt] coordinates {(0.25, 0.76080734) (0.50, 0.93321246) (0.75, 0.9979225) (1.0, 0.99977505)};
		\addplot[ultra thick, viryellow, mark=diamond*, mark size=1.5pt] coordinates {(0.25, 0.59265035) (0.50, 0.8057956) (0.75, 0.95391226) (1.0, 0.9828972)};
		
		\nextgroupplot[
		title={\color{black!10!gray}  BM},
		axis lines=left,
		xtick={0.25, 0.5, 0.75, 1},
		xticklabels={0.25, 0.5, 0.75, 1},
		ymin=0.1, ymax=1.01,
		]
		\addplot[ultra thick, virpurple, mark=square*, mark size=1.5pt] coordinates {(0.25, 0.90868485) (0.50, 0.9515885) (0.75, 0.98301923) (1.0, 0.9953362)};
		\addplot[ultra thick, virblue, mark=*, mark size=1.5pt] coordinates {(0.25, 0.7919229) (0.50, 0.9578477) (0.75, 0.9643815) (1.0, 0.97808534)};
		\addplot[ultra thick, virgreen, mark=triangle*, mark size=1.5pt] coordinates {(0.25, 0.5608772) (0.50, 0.8542218) (0.75, 0.93733895) (1.0, 0.9532735)};
		\addplot[ultra thick, viryellow, mark=diamond*, mark size=1.5pt] coordinates {(0.25, 0.33544135) (0.50, 0.53400004) (0.75, 0.8728872) (1.0, 0.92621654)};
		
		\nextgroupplot[
		title={\color{black!10!gray}  DMRG-I},
		axis lines=left,
		xtick={0.25, 0.5, 0.75, 1},
		xticklabels={0.25, 0.5, 0.75, 1},
		ymin=0.1, ymax=1.01,
		]
		\addplot[ultra thick, virpurple, mark=square*, mark size=1.5pt] coordinates {(0.25, 0.99991745) (0.50, 0.9999931) (0.75, 0.9999968) (1.0, 0.99999726)};
		\addplot[ultra thick, virblue, mark=*, mark size=1.5pt] coordinates {(0.25, 0.94496423) (0.50, 0.9999777) (0.75, 0.99999285) (1.0, 0.9999954)};
		\addplot[ultra thick, virgreen, mark=triangle*, mark size=1.5pt] coordinates {(0.25, 0.90189564) (0.50, 0.99707246) (0.75, 0.9994041) (1.0, 0.99988055)};
		\addplot[ultra thick, viryellow, mark=diamond*, mark size=1.5pt] coordinates {(0.25, 0.99442947) (0.50, 0.9913755) (0.75, 0.99653316) (1.0, 0.9981675)};
		
		\nextgroupplot[
		title={\color{black!10!gray} DMRG-II},
		axis lines=left,
		xtick={0.25, 0.5, 0.75, 1},
		xticklabels={0.25, 0.5, 0.75, 1},
		ymin=0.1, ymax=1.01,
		]
		\addplot[ultra thick, virpurple, mark=square*, mark size=1.5pt] coordinates {(0.25, 0.9952222) (0.50, 0.9999949) (0.75, 0.9999991) (1.0, 0.99999726)};
		\addplot[ultra thick, virblue, mark=*, mark size=1.5pt] coordinates {(0.25, 0.9368582) (0.50, 0.99968874) (0.75, 0.99999285) (1.0, 0.9999954)};
		\addplot[ultra thick, virgreen, mark=triangle*, mark size=1.5pt] coordinates {(0.25, 0.86013055) (0.50, 0.99849683) (0.75, 0.99970734) (1.0, 0.99997365)};
		\addplot[ultra thick, viryellow, mark=diamond*, mark size=1.5pt] coordinates {(0.25, 0.77826786) (0.50, 0.98845017) (0.75, 0.99532986) (1.0, 0.99717563)};

		
		\nextgroupplot[
		ylabel={Trace Distance},
		axis lines=left,
		xtick={0.25, 0.50, 0.75, 1.0},
		xticklabels={0.25, 0.50, 0.75, 1.0},
		ymin=1e-4, ymax=1,
		ymode=log,
		ytick={1e-4, 1e-3, 1e-2, 1e-1, 1},
		minor tick num=0,
		]
		\addplot[ultra thick, virpurple, mark=square*, mark size=1.5pt] coordinates {(0.25, 0.11421487) (0.50, 0.01088495) (0.75, 0.0015115) (1.0, 0.00110119)};
		\addplot[ultra thick, virblue, mark=*, mark size=1.5pt] coordinates {(0.25, 0.33465502) (0.50, 0.07334654) (0.75, 0.00933366) (1.0, 0.00159655)};
		\addplot[ultra thick, virgreen, mark=triangle*, mark size=1.5pt] coordinates {(0.25, 0.5510099) (0.50, 0.25567192) (0.75, 0.02074909) (1.0, 0.00299693)};
		\addplot[ultra thick, viryellow, mark=diamond*, mark size=1.5pt] coordinates {(0.25, 0.76833624) (0.50, 0.47203252) (0.75, 0.21190089) (1.0, 0.11167913)};
		
		\nextgroupplot[
		axis lines=left,
		xtick={0.25, 0.50, 0.75, 1.0},
		xticklabels={0.25, 0.50, 0.75, 1.0},
		ymin=1e-4, ymax=1,
		ymode=log,
		ytick={1e-4, 1e-3, 1e-2, 1e-1, 1},
		]
		\addplot[ultra thick, virpurple, mark=square*, mark size=1.5pt] coordinates {(0.25, 0.38202706) (0.50, 0.27636367) (0.75, 0.14897415) (1.0, 0.08126789)};
		\addplot[ultra thick, virblue, mark=*, mark size=1.5pt] coordinates {(0.25, 0.47707084) (0.50, 0.227704) (0.75, 0.18633078) (1.0, 0.148539)};
		\addplot[ultra thick, virgreen, mark=triangle*, mark size=1.5pt] coordinates {(0.25, 0.75719374) (0.50, 0.4727038) (0.75, 0.32582116) (1.0, 0.2695867)};
		\addplot[ultra thick, viryellow, mark=diamond*, mark size=1.5pt] coordinates {(0.25, 0.89523304) (0.50, 0.79016876) (0.75, 0.44326824) (1.0, 0.3399659)};
		
		\nextgroupplot[
		axis lines=left,
		xtick={0.25, 0.50, 0.75, 1.0},
		xticklabels={0.25, 0.50, 0.75, 1.0},
		ymin=1e-4, ymax=1,
		ymode=log,
		ytick={1e-4, 1e-3, 1e-2, 1e-1, 1},
		]
		\addplot[ultra thick, virpurple, mark=square*, mark size=1.5pt] coordinates {(0.25, 0.01225613) (0.50, 0.00510203) (0.75, 0.00158994) (1.0, 0.00137814)};
		\addplot[ultra thick, virblue, mark=*, mark size=1.5pt] coordinates {(0.25, 0.2455803) (0.50, 0.03334634) (0.75, 0.00597816) (1.0, 0.00470097)};
		\addplot[ultra thick, virgreen, mark=triangle*, mark size=1.5pt] coordinates {(0.25, 0.39748102) (0.50, 0.07171147) (0.75, 0.03258011) (1.0, 0.01475505)};
		\addplot[ultra thick, viryellow, mark=diamond*, mark size=1.5pt] coordinates {(0.25, 0.4368726) (0.50, 0.12305336) (0.75, 0.07994174) (1.0, 0.05728783)};
		
		\nextgroupplot[
		axis lines=left,
		xtick={0.25, 0.50, 0.75, 1.0},
		xticklabels={0.25, 0.50, 0.75, 1.0},
		ymin=1e-4, ymax=1,
		ymode=log,
		ytick={1e-4,1e-3, 1e-2, 1e-1, 1},
		]
		\addplot[ultra thick, virpurple, mark=square*, mark size=1.5pt] coordinates {(0.25, 0.09080536) (0.50, 0.00301315) (0.75, 0.0012125) (1.0, 0.00091314)};
		\addplot[ultra thick, virblue, mark=*, mark size=1.5pt] coordinates {(0.25, 0.26008105) (0.50, 0.0199744) (0.75, 0.00472261) (1.0, 0.00431881)};
		\addplot[ultra thick, virgreen, mark=triangle*, mark size=1.5pt] coordinates {(0.25, 0.47801507) (0.50, 0.05189832) (0.75, 0.0227311) (1.0, 0.00664242)};
		\addplot[ultra thick, viryellow, mark=diamond*, mark size=1.5pt] coordinates {(0.25, 0.59182006) (0.50, 0.13953146) (0.75, 0.09460422) (1.0, 0.07299805)};

		
		\nextgroupplot[
		ylabel={Time (s)},
		axis lines=left,
		ymode=log,
		ymin=1e-1, ymax=1e3,
		xtick={0.25, 0.50, 0.75, 1.0},
		xticklabels={0.25, 0.50, 0.75, 1.0},
		xlabel={$\alpha$},
		ytick={1e-1, 1e0, 1e1, 1e2, 1e3},
		]
		\addplot[ultra thick, virpurple, mark=square*, mark size=1.5pt] coordinates {(0.25, 1.6490049) (0.50, 1.8191973) (0.75, 1.8542813) (1.0, 1.8624341)};
		\addplot[ultra thick, virblue, mark=*, mark size=1.5pt] coordinates {(0.25, 7.299287) (0.50, 7.4553275) (0.75, 7.843272) (1.0, 7.992088)};
		\addplot[ultra thick, virgreen, mark=triangle*, mark size=1.5pt] coordinates {(0.25, 31.218777) (0.50, 32.71353) (0.75, 36.101185) (1.0, 41.007145)};
		\addplot[ultra thick, viryellow, mark=diamond*, mark size=1.5pt] coordinates {(0.25, 155.33704) (0.50, 145.54695) (0.75, 159.77586) (1.0, 173.72462)};
		
		\nextgroupplot[
		axis lines=left,
		ymode=log,
		ymin=1e-1, ymax=1e3,
		xtick={0.25, 0.50, 0.75, 1.0},
		xticklabels={0.25, 0.50, 0.75, 1.0},
		xlabel={$\alpha$},
		ytick={1e-1, 1e0, 1e1, 1e2, 1e3},
		]
		\addplot[ultra thick, virpurple, mark=square*, mark size=1.5pt] coordinates {(0.25, 0.41501987) (0.50, 0.46687663) (0.75, 0.5197587) (1.0, 0.59371805)};
		\addplot[ultra thick, virblue, mark=*, mark size=1.5pt] coordinates {(0.25, 0.9049369) (0.50, 1.4454622) (0.75, 1.951388) (1.0, 2.6752672)};
		\addplot[ultra thick, virgreen, mark=triangle*, mark size=1.5pt] coordinates {(0.25, 3.92689) (0.50, 7.0587287) (0.75, 10.317442) (1.0, 13.519121)};
		\addplot[ultra thick, viryellow, mark=diamond*, mark size=1.5pt] coordinates {(0.25, 18.618174) (0.50, 36.027283) (0.75, 58.102005) (1.0, 76.02945)};
		
		\nextgroupplot[
		axis lines=left,
		ymode=log,
		ymin=1e-1, ymax=1e3,
		xtick={0.25, 0.50, 0.75, 1.0},
		xticklabels={0.25, 0.50, 0.75, 1.0},
		xlabel={$\alpha$},
		ytick={1e-1, 1e0, 1e1, 1e2, 1e3},
		]
		\addplot[ultra thick, virpurple, mark=square*, mark size=1.5pt] coordinates {(0.25, 0.18427801) (0.50, 0.17833042) (0.75, 0.18054366) (1.0, 0.16510582)};
		\addplot[ultra thick, virblue, mark=*, mark size=1.5pt] coordinates {(0.25, 0.6243304) (0.50, 0.3345927) (0.75, 0.37780297) (1.0, 0.41195714)};
		\addplot[ultra thick, virgreen, mark=triangle*, mark size=1.5pt] coordinates {(0.25, 1.1077745) (0.50, 0.72117555) (0.75, 0.80233276) (1.0, 0.903525)};
		\addplot[ultra thick, viryellow, mark=diamond*, mark size=1.5pt] coordinates {(0.25, 1.5433831) (0.50, 1.2564029) (0.75, 1.5026537) (1.0, 1.8208897)};
		
		\nextgroupplot[
		axis lines=left,
		ymode=log,
		ymin=1e-1, ymax=1e3,
		xtick={0.25, 0.50, 0.75, 1.0},
		xticklabels={0.25, 0.50, 0.75, 1.0},
		xlabel={$\alpha$},
		ytick={1e-1, 1e0, 1e1, 1e2, 1e3},
		]
		\addplot[ultra thick, virpurple, mark=square*, mark size=1.5pt] coordinates {(0.25, 0.1060617) (0.50, 0.1246134) (0.75, 0.12456405) (1.0, 0.11818743)};
		\addplot[ultra thick, virblue, mark=*, mark size=1.5pt] coordinates {(0.25, 0.4140135) (0.50, 0.23796439) (0.75, 0.33322394) (1.0, 0.30116856)};
		\addplot[ultra thick, virgreen, mark=triangle*, mark size=1.5pt] coordinates {(0.25, 1.1159481) (0.50, 0.50760484) (0.75, 0.54292464) (1.0, 0.63420093)};
		\addplot[ultra thick, viryellow, mark=diamond*, mark size=1.5pt] coordinates {(0.25, 1.4205725) (0.50, 1.609776) (0.75, 2.0404491) (1.0, 1.9977233)};
		
		\addlegendentry{$N=4$}
		\addlegendentry{ $N=5$}
		\addlegendentry{$N=6$}
		\addlegendentry{$N=7$}
		
	\end{groupplot}

\end{tikzpicture}

%% file: figures/loss_sweep.tex
\begin{tikzpicture}[scale=0.7]
	\begin{axis}[
		width=\linewidth,
		height=0.4\linewidth,
		axis lines=left,
        clip=false,
		grid=both,
		ymode=log,
		xmin=0, xmax=21,
		ymin=1e-6, ymax=10,
		xlabel={Active site},
		ylabel={Loss},
		xtick={0, 1, 2, 3, 4, 5, 6, 7, 8, 9, 10, 11, 12, 13, 14, 15, 16, 17, 18, 19, 20, 21},
		xticklabels={0, 1, 2, 3, 4, 5, 4, 3, 2, 1, 0, 1, 2, 3, 4, 5, 4, 3, 2, 1, 0, 1},
		ytick={1e-6, 1e-5, 1e-4, 1e-3, 1e-2, 1e-1, 1e1},
		every axis x label/.append style={
			at={(axis description cs:0.5,-0.2)},
			anchor=north, color=virpurple!60!gray
		},
		every axis y label/.append style={
			at={(axis description cs:-0.12,0.5)},
			anchor=south, color=black!10!gray
		},
		x axis line style={yshift=-8pt,-,color=gray},
		y axis line style={xshift=-8pt,-,color=gray},
		xticklabel style={yshift=-8pt,color=virpurple, font=\normalsize, rotate=90},
		yticklabel style={xshift=-8pt,color=gray, font=\normalsize},
		xtick style={yshift=-8pt, line width=0.3pt},
		ytick style={xshift=-8pt, line width=0.3pt},
		xtick align=inside,
		ytick align=inside,
		extra x ticks={0.5, 1.5, 2.5, 3.5, 4.5, 5.5, 6.5, 7.5, 8.5},
		extra x tick labels={(0,1), (1,2), (2,3), (3,4), (4,5), (3,4), (2,3), (1,2), (0,1), (1,2)},
		extra x tick style={
			ticklabel pos=top,
			xticklabel style={
            font=\normalsize,
            color=virgreen,
            rotate=0,
            xshift=20pt,
            yshift=-8pt,
            anchor=south
            }
		},
		legend style={
			at={(0.5,-0.35)},
			anchor=north,
			legend columns=2,
			draw=none,
			fill=none,
			text=black!10!gray,
			font=\normalsize,
			/tikz/every even column/.append style={column sep=0.2cm}
		},
		legend image post style={line width=2pt}
	]

	\addplot[
		ultra thick,
		virpurple,
		mark=square*,
		mark size=2pt
	] coordinates {
		(0, 3.6196680e+00)
		(1, 3.5491161e+00)
		(2, 3.3599801e+00)
		(3, 3.2274308e+00)
		(4, 2.4873309e+00)
		(5, 2.0676384e+00)
		(6, 7.9504889e-01)
		(7, 6.6644800e-01)
		(8, 5.4464698e-01)
		(9, 4.4580698e-01)
		(10, 2.2737072e-01)
		(11, 7.1149662e-02)
        (12, 7.1149640e-02)
		(13, 7.1149632e-02)
		(14, 7.0232265e-02)
		(15, 3.2003820e-02)
		(16, 5.8076726e-03)
		(17, 5.8073839e-03)
		(18, 5.8073839e-03)
		(19, 5.8073858e-03)
        (20, 5.8073867e-03)
        (21, 4.2008946e-06)        
	};
	\addlegendentry{DMRG-1}

	\addplot[
		ultra thick,
		virgreen,
		mark=triangle*,
		mark size=2pt,
	] coordinates {
		(0.5, 3.5491161e+00)
		(1.5, 3.3444414e+00)
		(2.5, 3.1886697e+00)
		(3.5, 2.4055896e+00)
		(4.5, 1.8383833e+00)
		(5.5, 2.3771670e-01)
		(6.5, 2.2443798e-01)
		(7.5, 1.6293150e-01)
		(8.5, 4.0793748e-06)
	};
	\addlegendentry{DMRG-2}

	\node[
		font=\normalsize,
		anchor=south
	] at (axis description cs:0.5,1.25) {\textcolor{virgreen}{Active site pair}};

	\end{axis}

	\begin{axis}[
		at={(13cm,2cm)},
		width=4cm,
		height=3cm,
		xlabel={Half sweep},
		ylabel={$R_{\mathrm{max}}$\\(median)},
		ylabel style={align=center, font=\small},
		xlabel style={font=\small},
		ticklabel style={font=\small},
        xtick={1,2,3,4},
        xticklabels={1,2,3,4},
		ymin=0, ymax=12,
		xmin=0.5, xmax=4.5,
		ybar=0pt,
		bar width=5pt,
		grid=both,
		grid style={line width=.1pt, draw=gray!20},
		axis background/.style={fill=white}
	]
		\draw[dotted, thick, black] (axis cs: 0.2, 10) -- (axis cs: 4.2, 10);

		\addplot[fill=virpurple, draw=none] coordinates {
			(1, 2)
			(2, 4)
			(3, 8)
			(4, 10)
		};

		\addplot[fill=virgreen, draw=none] coordinates {
			(1, 4)
			(2, 10)
		};
	\end{axis}
\end{tikzpicture}

%% file: figures/scalability.tex
\begin{tikzpicture}[scale=0.71]
\begin{groupplot}[
    group style={
        group size=3 by 1,
        horizontal sep=2.3cm
    },
    width=0.4\textwidth,
  	legend style={
			at={(-1,-0.6)},
			anchor=north,
			legend columns=2,
			draw=none,
			fill=none,
			text=black!10!gray,
			font=\normalsize,
			/tikz/every even column/.append style={column sep=0.2cm}
		},
	legend image post style={line width=2pt},
    grid=both,
]

\nextgroupplot[
    xlabel={$N$},
    ylabel={\Large $\frac{\|\rho-\hat{\rho}\|_F}{\|\rho\|_F}$},
    xtick={4,6,8,10,12},
    ymin=1e-4, ymax=1e0,
	ymode=log,
	ytick={1e-4, 1e-3, 1e-2, 1e-1, 1e0},
    axis lines=middle,
    axis x line=bottom,
    axis y line=left,
    enlargelimits=true,
    every axis x label/.append style={
        at={(axis description cs:0.5,-0.35)},
        anchor=north,
        color=gray
    },
    every axis y label/.append style={
        at={(axis description cs:-0.35,0.5)},
        anchor=south,
        rotate=90,
        color=gray
    },
    x axis line style={yshift=-8pt, -, color=gray},
    y axis line style={xshift=-8pt, -, color=gray},
    xticklabel style={yshift=-8pt, color=gray, font=\normalsize },
    yticklabel style={xshift=-8pt, color=gray, font=\normalsize},
    xtick style={yshift=-8pt, line width=0.2pt},
    ytick style={xshift=-8pt, line width=0.2pt},
    xtick align=inside,
    ytick align=inside,
]

\addplot[
    ultra thick,
    virpurple,
    mark=square*,
    mark size=2pt
] coordinates {
    (4,0.00041414)
    (6,0.01375016)
    (8,0.04298998)
    (10,0.09395632)
    (12,0.20909882)
};

\addplot[
    ultra thick,
    virgreen,
    mark=triangle*,
    mark size=2pt
] coordinates {
    (4,0.00043453)
    (6,0.00071564)
    (8,0.03841372)
    (10,0.1089289)
    (12,0.18638715)
};

\nextgroupplot[
    xlabel={$N$},
    ylabel={\Large
        $\frac{\|\mathbf{y}_{\mathrm{test}}
        -\hat{\mathbf{y}}_{\mathrm{test}}\|_2}
        {\|\mathbf{y}_{\mathrm{test}}\|_2}$
    },
    xtick={4,6,8,10,12},
    ymin=1e-4, ymax=1e-1,
	ymode=log,
	ytick={1e-4, 1e-3,  1e-2, 1e-1},
    axis lines=middle,
    axis x line=bottom,
    axis y line=left,
    enlargelimits=true,
    every axis x label/.append style={
        at={(axis description cs:0.5,-0.35)},
        anchor=north,
        color=gray
    },
    every axis y label/.append style={
        at={(axis description cs:-0.3,0.5)},
        anchor=south,
        rotate=90,
        color=gray
    },
    x axis line style={yshift=-8pt, -, color=gray},
    y axis line style={xshift=-8pt, -, color=gray},
    xticklabel style={yshift=-8pt, color=gray, font=\normalsize},
    yticklabel style={xshift=-8pt, color=gray, font=\normalsize},
    xtick style={yshift=-8pt, line width=0.2pt},
    ytick style={xshift=-8pt, line width=0.2pt},
    xtick align=inside,
    ytick align=inside,
]

\addplot[
    ultra thick,
    virpurple,
    mark=square*,
    mark size=2pt
] coordinates {
    (4,0.00021636)
    (6,0.00522923)
    (8,0.0129264)
    (10,0.0395738)
    (12,0.08936806)
};

\addplot[
    ultra thick,
    virgreen,
    mark=triangle*,
    mark size=2pt
] coordinates {
    (4,0.00022244)
    (6,0.00026332)
    (8,0.01275815)
    (10,0.04261997)
    (12,0.08162155)
};

\nextgroupplot[
    xlabel={$N$},
    ylabel={\textcolor{gray}{Time (s)}},
    xtick={4,6,8,10,12},
    ymin=1e-1, ymax=1e4,
	ymode=log,
	ytick={1e-1, 1e1, 1e2, 1e3, 1e4},
    axis lines=middle,
    axis x line=bottom,
    axis y line=left,
    enlargelimits=true,
    every axis x label/.append style={
        at={(axis description cs:0.5,-0.35)},
        anchor=north,
        color=gray
    },
    every axis y label/.append style={
        at={(axis description cs:-0.3,0.5)},
        anchor=south,
        rotate=90,
        color=gray
    },
    x axis line style={yshift=-8pt, -, color=gray},
    y axis line style={xshift=-8pt, -, color=gray},
    xticklabel style={yshift=-8pt, color=gray, font=\normalsize},
    yticklabel style={xshift=-8pt, color=gray, font=\normalsize},
    xtick style={yshift=-8pt, line width=0.2pt},
    ytick style={xshift=-8pt, line width=0.2pt},
    xtick align=inside,
    ytick align=inside,
]

\addplot[
    ultra thick,
    virpurple,
    mark=square*,
    mark size=2pt
] coordinates {
    (4,0.3738035)
    (6,2.1319857)
    (8,15.351702)
    (10,49.93495)
    (12,119.13919)
};
\addlegendentry{DMRG-I}

\addplot[
    ultra thick,
    virgreen,
    mark=triangle*,
    mark size=2pt
] coordinates {
    (4,2.5314415e-01)
    (6,1.9890008e+00)
    (8,3.5652290e+01)
    (10,1.7181158e+02)
    (12,5.5851709e+02)
};
\addlegendentry{DMRG-II}

\end{groupplot}

\end{tikzpicture}

%% file: figures/K_approx.tex
\begin{tikzpicture}[scale=0.75]
  \begin{groupplot}[
    group style={group size=2 by 1, horizontal sep=2cm},
    width=0.4\textwidth,
	legend style={
			at={(-0.2,-0.45)},
			anchor=north,
			legend columns=4,
			draw=none,
			fill=none,
			text=black!10!gray,
			font=\normalsize,
			/tikz/every even column/.append style={column sep=0.2cm}
		},
	legend image post style={line width=2pt}
  ]
    \nextgroupplot[
    xlabel={$N$},
    ylabel={Fidelity},
    axis lines=middle,
    enlargelimits=true,
    axis x line = bottom,
    axis y line =left,
    grid=both,
    grid style={line width=0.2pt, draw=gray!70},
    xtick={4, 5, 6, 7, 8},
	xticklabels={4, 5, 6, 7, 8},
    xmin=4, xmax=8,
	ymin=0.6, ymax=1.01,
    scale=1,
    every axis x label/.append style={
    at={(axis description cs:0.5,-0.3)},
    anchor=north, color=gray
    },
  every axis y label/.append style={
    at={(axis description cs:-0.3,0.5)},
    anchor=south,
    rotate=90, color=gray},
    x axis line style={yshift=-10pt,-,color=gray},
    xticklabel style={yshift=-10pt,color=gray},
    y axis line style={xshift=-10pt,-,color=gray},
    yticklabel style={xshift=-10pt,color=gray},
    xtick style={yshift=-10pt, line width=0.5pt},
    ytick style={xshift=-10pt, line width=0.5pt},
    xtick align=inside,
    ytick align=inside,
    ]
      
    \addplot[ultra thick, virpurple, mark=square*, mark size=2pt] coordinates {(4, 0.78104992) (5, 0.84061419) (6,0.72126958) (7, 0.68082803) (8, 0.69462742)};
      
     \addplot[ultra thick, virblue, mark=*, mark size=2pt] coordinates {(4, 0.99999889) (5, 0.99999606) (6, 0.9999448) (7, 0.99684544) (8, 0.9887497)};
      
      \addplot[ultra thick, virgreen, mark=triangle*, mark size=2pt] coordinates {(4, 0.99999901) (5, 0.99999564) (6, 0.98995675) (7, 0.97979557) (8, 0.9667963)};

      \addplot[ultra thick, viryellow, mark=diamond*, mark size=2pt] coordinates {(4, 0.99999907) (5, 0.99999612) (6, 0.99187798) (7, 0.97438681) (8, 0.96825101)};
      
    \nextgroupplot[
      xlabel={$N$},
      xtick={4, 5, 6, 7, 8},
	   xticklabels={4, 5, 6, 7, 8},
      xmin=4, xmax=8,
      ylabel={\textcolor{gray}{Time (s)}},
      axis lines=middle,
      enlargelimits=true,
      axis x line = bottom,
      axis y line = left,
      ymode=log,
      ymin=0.1,
      ymax=100,
       grid=both,
      grid style={line width=0.2pt, draw=gray!70},
       every axis x label/.append style={
    	at={(axis description cs:0.5,-0.3)},
    	anchor=north, color=gray
    },
    every axis y label/.append style={
    	at={(axis description cs:-0.3,0.5)},
    	anchor=south,
    	rotate=90, color=gray},
    x axis line style={yshift=-10pt,-,color=gray},
    xticklabel style={yshift=-10pt,color=gray},
    y axis line style={xshift=-10pt,-,color=gray},
    yticklabel style={xshift=-10pt,color=gray},
    xtick style={yshift=-10pt, line width=0.5pt},
    ytick style={xshift=-10pt, line width=0.5pt},
    xtick align=inside,
    ytick align=inside,
    ]

    \addplot[ultra thick, virpurple, mark=square*, mark size=2pt] coordinates {(4, 0.42406797) (5, 0.79139173) (6, 2.23458004) (7, 7.86047089) (8, 23.44440258)};
\addlegendentry{$K=1$}

\addplot[ultra thick, virblue, mark=*, mark size=2pt] coordinates {(4, 0.15306652) (5, 0.33952713) (6, 0.74188578) (7, 3.59607601) (8, 14.0734818)};
\addlegendentry{$K=2$}

\addplot[ultra thick, virgreen, mark=triangle*, mark size=2pt] coordinates {(4, 0.15985775) (5, 0.36777747) (6, 0.77512062) (7, 3.84525013) (8, 12.95697355)};
\addlegendentry{$K=3$}

\addplot[ultra thick, viryellow, mark=diamond*, mark size=2pt] coordinates {(4, 0.14876652) (5, 0.37134099) (6, 0.7729857) (7, 3.78074789) (8, 12.37205839)};
\addlegendentry{$K=4$}

  \end{groupplot}

\end{tikzpicture}